\documentclass[]{jfm}

\usepackage{graphicx}
\usepackage{newtxtext}
\usepackage{newtxmath}
\usepackage{natbib}
\usepackage{hyperref}
\usepackage{textcomp}
\usepackage{bm}
\usepackage{amsmath}
\usepackage{comment}
\hypersetup{
    colorlinks = true,
    urlcolor   = blue,
    citecolor  = black,
}

\newcommand{\RomanNumeralCaps}[1]

\newcommand*{\reviewerA}[1]{\textcolor{black}{#1}}
\newcommand*{\add}[1]{\textcolor{black}{#1}}

\title{Unsteady large-scale wake structure behind\\levitated freestream-aligned circular cylinder}

\author{Sho Yokota\aff{1}
  \corresp{\email{sho.yokota.r1@dc.tohoku.ac.jp}}
 \and Taku Nonomura\aff{1}}

\affiliation{\aff{1}Department of Aerospace Engineering, Graduate School of Engineering, Tohoku University,\\Sendai, Miyagi, 980-8579, Japan}

\begin{document}
\maketitle

\begin{abstract}
The relationships between characteristic large-scale wake structures appearing behind a freestream-aligned circular cylinder are investigated and discussed from the velocity field obtained by wind tunnel tests. The tests were conducted under a supportless condition using a magnetic suspension and balance system and stereo PIV measurements at a Reynolds number of $3.46\times10^4$. The velocity fields were analysed with a modal decomposition combining azimuthal Fourier decomposition and proper orthogonal decomposition. The wake behind the freestream-aligned circular cylinder with three different fineness ratios of 1.0, 1.5 and 2.0 was investigated, and the wake structures in a nonreattaching flow formed by the cylinder at the fineness ratio of 1.0 are mainly discussed in the present study. Four characteristic large-scale wake structures of recirculation bubble pumping, azimuthal shear mode, large-scale vortex shedding and streaks are identified and mainly focused on in the present study. The state of the vortex shedding is classified into three; anticlockwise/clockwise circular and flapping patterns. Each state has a relationship with the azimuthal shear mode and it tends to appear when the state is circular. Furthermore, from the analysis of the relationship between modes, the recirculation bubble pumping is found to be related to the vortex shedding position in the radial direction and the strength of the streaks. Particularly, analysis of causality shows that the recirculation bubble pumping is affected by them in the low-frequency range.
\end{abstract}

\begin{keywords}
Authors should not enter keywords on the manuscript, as these must be chosen by the author during the online submission process and will then be added during the typesetting process (see \href{https://www.cambridge.org/core/journals/journal-of-fluid-mechanics/information/list-of-keywords}{Keyword PDF} for the full list).  Other classifications will be added at the same time.
\end{keywords}


\section{Introduction}\label{sec:Intro}
The flow around bluff bodies is often found in our surroundings and in industrial fields. Typical examples are aircraft gears, railroad vehicles, and buildings such as skyscrapers and bridge piers. Comprehension of the aerodynamic characteristics of these blunt-head applications is essential for evaluating their impact on the economy, safety, and the living environment, such as noise. Representative bluff bodies, such as circular cylinders, rectangular prisms, and spheres, have been investigated by many researchers \citep{berger1990coherent, nakaguchi1968experimental}. On the other hand, there are relatively few studies on a freestream-aligned circular cylinder, which is one of the bluff bodies. A freestream-aligned circular cylinder is a cylinder in which the central axis is parallel to the direction of the freestream. Applications with shapes similar to the freestream-aligned circular cylinder include oil tanks, engine canisters \citep{prosser2016numerical}, re-entry capsules\citep{ohmichi2019numerical}, and automobile door mirrors \citep{yang2015low}. In addition, an axisymmetric bluff body with a nose \citep{rigas2014low, rigas2015diffusive, zhang2023coherent}, a three-dimensional rectangular cylinder \citep{greenwell2011modelling}, and the Ahmed model \citep{grandemange2013turbulent}, which is similar to the freestream-aligned circular cylinder, have been studied.

Studies on the freestream-aligned circular cylinder have been conducted both numerically \citep{yang2014numerical, yang2015low, tian2016large, tian2017direct, gao2018flow, prosser2016numerical, chongsiripinyo2020decay, nidhan2020spectral} and experimentally \citep{johansson2006far1, johansson2006far2, higuchi2006axial, higuchi2008sting, bobinski2014instabilities, shinji2020aerodynamic, kuwata2021flow, yokota2021analysis, yokota2022instability, yokota2023effect}. The flow around the freestream-aligned circular cylinder is classified into two main types based on the time-averaged velocity field. One type is nonreattaching flow in which the separated flow at the leading edge does not reattach on the curved surface, and it appears when the fineness ratio $L/D$ ($L$: length, $D$: diameter) is less than 1.5. The other is reattaching flow, in which the separated flow reattaches on the curved surface, and is observed when $L/D$ is greater than 1.5. The previous studies have mainly focused on flows around the cylinder at $L/D=0$ or close to 0, from low Reynolds numbers where laminar flows are formed \citep{fabre2008bifurcations, bobinski2014instabilities, tian2017direct, gao2018flow} to Reynolds numbers where fully developed turbulence appears \citep{berger1990coherent, johansson2006far1, johansson2006far2, yang2014numerical, yang2015low, tian2016large, chongsiripinyo2020decay, nidhan2020spectral}. \reviewerA{A steady state (SS) mode} with longitudinal vortex pairs, a reflectional symmetry breaking (RSB) mode, and a standing wave (SW) mode were observed in the low Reynolds number region \citep{fabre2008bifurcations}. The RSB mode has also been observed in the investigation of disk wake at $Re=10^4$, where it is known that the streamwise vorticity exhibits a ``Yin-Yang'' pattern in the plane perpendicular to the freestream \citep{yang2014numerical}. This ``Yin-Yang'' pattern due to the RSB mode was also observed in the wake of an axisymmetric bluff body forming an attached flow \citep{zhang2023coherent}. Furthermore, a reflectional symmetry preserving (RSP) mode was observed by \citet{rigas2014low} and \citet{pavia2019three} in the wake of an axisymmetric bluff body. On the other hand, the SS mode is not observed after the critical Reynolds number at which the transition to the RSB mode occurs, and the wake is reported as a symmetry-broken flow. This symmetry-broken flow was also observed in the ground-mounted Ahmed model wake \citep{grandemange2013turbulent}. The appearance of symmetry-broken modes of RSP and RSB modes in the wake of the freestream-aligned circular cylinder is expected, but has not been discussed in the studies on the turbulent wake of the cylinder at $L/D\ge0.3$ \citep{prosser2016numerical, nonomura2018effect, shinji2020aerodynamic, yokota2021analysis, kuwata2021flow}. The relationship between the characteristic flow structures occurring in the wake and the aerodynamic forces acting on the cylinder has been well discussed in their studies.

Three characteristic flow structures have been confirmed in the wake of the cylinder from the velocity and pressure fluctuation spectra: recirculating bubble pumping ($St < 0.05$), large-scale vortex shedding ($St \approx 0.13$), and the Kelvin--Helmholtz instability. Here, the Strouhal number is a nondimensional frequency defined by $St = fD/U$ ($f$: frequency, $U$: freestream velocity). The first two phenomena relate to aerodynamic force fluctuations acting on the cylinder. Firstly, the recirculation bubble pumping is a phenomenon of axisymmetric fluctuations in which the size of the recirculation region formed behind the cylinder expands or contracts in the freestream direction and is more noticeable in the case of the nonreattaching flow. Simultaneously, the pressure field at the base of the cylinder fluctuates, which appears as a drag force fluctuation acting on the cylinder. In contrast, the PSD distribution shown by \citet{yokota2021analysis} indicates that the amplitude of the fluctuation due to the recirculation bubble pumping is smaller for the reattaching flow at $L/D\ge1.5$ compared to the case of the nonreattaching flow. Moreover, \citet{nonomura2018effect} confirmed that the fluctuations in the low-frequency region around $f=1$~Hz ($St=0.02$) were smaller only for $L/D=2.0$ from the PSD of the pressure fluctuations at the centre of the cylinder base for $L/D=0.75\text{--}2.0$. Secondly, large-scale vortex shedding is a phenomenon of antisymmetric fluctuations in which a large vortex structure containing small vortices is shed in the lateral direction from the downstream end of the recirculation region and is more apparent for $L/D\le1.5$. Fluctuations due to this flow structure are dominant in the flow around the freestream-aligned circular cylinder and appear as pressure fluctuations not only downstream of the recirculation region but also near the sides of the cylinder. The result appears as lateral force fluctuations acting on the cylinder. Relatively large fluctuations around the fluctuation frequency of the large-scale vortex shedding were also observed in the PSD for $L/D=2.0$, which was not mentioned by \citet{yokota2021analysis}. Note that the fluctuation level is smaller than in the case of $L/D\le1.5$. From the above, although the relationship between the phenomena and aerodynamic fluctuations and the $L/D$ dependency on the fluctuations by the characteristic phenomena have been well discussed in previous studies, the interrelationships between the phenomena have not been fully investigated.

An interesting interrelationship between the phenomena is the relationship between the recirculating bubble pumping and fluctuations in the shedding position of large-scale vortex structures, as suggested by \citet{yang2015low}. They investigated numerically the flow around a disc ($L/D = 0.2$), in which they focused on the fluctuations of the shedding position of the large-scale vortex structure and showed that the switching of the rotational direction of the shedding position could be related to fluctuations in the recirculating bubble pumping. However, coherence and phase differences calculated from the velocity fluctuations due to the bubble pumping and positional fluctuations of vortex shedding have not been reported. No physical discussion has yet taken place on this point, and the mechanism of the switching is not clear. Numerical simulations provide the three velocity components over a three-dimensional field, whereas obtaining a long-duration flow field is difficult from the point of view of computational resources. In particular,  it is necessary to obtain data containing sufficient periods of phenomena with low-frequency fluctuations such as the recirculation bubble pumping, for comprehension of the aforementioned relationships. This requires the experimental acquisition of velocity fields with suitable time resolution and sufficient data length.

Investigations of the flow around three-dimensional objects have been carried out both numerically and experimentally. In most cases of experiments, the model is fixed in the channel by means of support. However, problems arise when stings, struts, wires, etc. used as supports interfere with the flow, altering the large-scale wake structure \citep{tashiro2023experimental}. This is known as support interference and makes it difficult to comprehend the actual large-scale wake structure and aerodynamic characteristics that occur in the flow around an object. The main experimental methods by which support interference can be eliminated are free-fall tests \citep{zhong2011experimental}, ballistic flight tests \citep{may1953free} and wind tunnel tests using a magnetic support balance system (MSBS) \citep{higuchi2006axial}. Among these methods, the MSBS, which allows steady model support, is suitable for the investigation of flows around objects. The MSBS is a device that can levitate and support models through the interaction between the magnetic field produced by the coil system and the permanent magnets inside the model.

Several wind tunnel tests using the MSBS have been carried out on freestream-aligned circular cylinders. In recent years, experiments have also combined particle image velocimetry (PIV). The velocity measurement planes in these previous studies can be divided into two main categories. The first is the case where the measurement plane is set parallel to the freestream through the cylinder axis, which has been adopted by \cite{higuchi2008sting}, \cite{yokota2021analysis, yokota2022instability, yokota2023effect} and \cite{kuwata2021flow}. This plane can capture fluctuations due to recirculating bubble pumping, but obviously not the azimuthal fluctuations in the position of the large-scale vortex structure. The second is the case where the measurement plane is perpendicular to the freestream and was adopted in the study by \cite{higuchi2008sting}. They obtained non-time-resolved two-dimensional and two-component velocity data that could not be used for frequency analysis, although the vortex emission position could be captured. Furthermore, the relationship between the bubble pumping and vortex shedding position has not been discussed because the velocity in the freestream direction has not been measured. Thus, the measurement plane perpendicular to the freestream with a two-dimensional, three-component PIV should enable the discussion of unexplained relationships between phenomena.

The objective of the present study is to clarify the three-dimensional large-scale wake structure formed behind a freestream-aligned circular cylinder. Wind tunnel tests under support interference-free conditions were conducted using the MSBS and the stereo PIV measurement system newly developed and installed. Discussion between phenomena is provided mainly from the results of mode decomposition for the measured velocity data. Our group has discussed the phenomena in the wake by adapting two-dimensional two-component velocity field measurements in a plane parallel to the freestream and a pressure measurement system using wireless transmission to experiments with the MSBS \citep{nonomura2018effect, shinji2020aerodynamic, yokota2021analysis, kuwata2021flow}. The present study is the first experience of measuring a two-dimensional three-component velocity field in an interference-free condition, and the results of the velocity field in a plane perpendicular to the freestream are considered to improve our understanding of the wake phenomena. In particular, the present study can clarify the shedding pattern of the large-scale vortex shedding which was unexplained in the previous studies.

\section{Experimental apparatus}\label{sec:Setup}
\subsection{Model}\label{sec:Model}
\begin{figure}
    \centering
    \includegraphics[width=13cm]{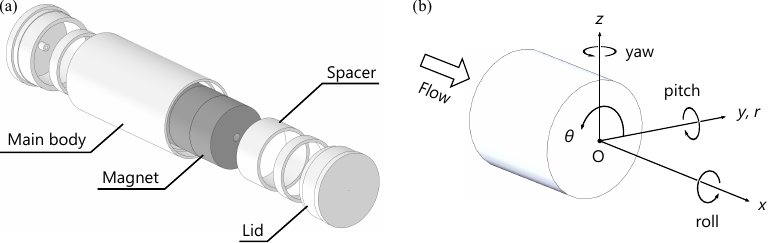}
    \caption{\label{fig:model} (a) Schematic of the cylindrical model and (b) the coordinate system in the present study}
\end{figure}

The flow around a freestream-aligned circular cylinder is divided into the nonreattaching flow and reattaching flow approximately at $L/D = 1.5$. Cylindrical models with $L/D =$ 1.0, 1.5 and 2.0 were used for wind tunnel tests in the present study, and both flows were investigated. A schematic of the model is shown in Fig.\ref{fig:model} (a). The model was made by machining polyoxymethylene, and the arithmetic mean roughness of the model surface and the corner edge were 1.18~\textmu m and R0.136, respectively. The model diameter is 50 mm regardless of $L/D$. The blockage rate in the wind tunnel test was 2.2\%. Permanent magnets required for magnetic levitation and support by MSBS are inserted inside the model. The inserted permanent magnets were cylindrical with an outer diameter of 40 mm, an inner diameter of 5 mm and a length of 20 mm, with two or three magnets connected lengthwise according to $L/D$. The outside of the model is basically white. Still, there is a black band for measuring the model position by the sensor subsystem of the MSBS, and the base of the model is painted black to suppress reflections of the laser light for PIV measurement.

The Cartesian coordinate system in the present study is based on the cylindrical model with an angle of attack to the freestream of 0~deg, as shown in Fig.~\ref{fig:model} (b). The centre of the base of the model is the origin, the $x$ axis is set in the freestream direction corresponding to the cylinder axis, the $z$ axis is set vertically upwards and the $y$ axis is set to form a right-handed system. The pitch $\theta$, yaw $\psi$ and roll $\phi$ angles are defined around the $y$, $z$ and $x$ axes respectively. Furthermore, a cylindrical coordinate system is defined. The $x$ axis is the same as in the Cartesian coordinate system, but the $r$ axis is the axis perpendicular to the circumference of the model from the origin, and the $\theta$ axis is the axis that is positive anticlockwise when the cylinder is viewed from behind. Here, the positive part of the $y$ axis is $\theta=0$~deg.

\subsection{Wind tunnel}\label{sec:WT}
Tohoku University--Basic Aerodynamics Research Tunnel (T-BART) was used in the tests. This wind tunnel is a suction-type wind tunnel with a closed test section. The dimensions of the cross section of the test section are usually 300~mm~$\times$~300~mm square, whereas the test section for stereo PIV used in the present study has a cross section of 296~mm~$\times$~300~mm. The reason for the smaller vertical size was to build a mirror system into the measuring section and the laser sheet was irradiated in a plane perpendicular to the freestream. The mirror systems are discussed in {\S}~\ref{sec:PIV}. The freestream velocity in T-BART can be set in the range of 5--60~m/s, at which the degree of turbulence is less than 0.5\%. Refer to appendix \ref{sec:FSVel} for the effects of changing the cross-sectional dimension of the test section. The freestream velocity $U$ was set to 10.5~m/s in the experiment, which corresponds to a Reynolds number of $3.46\times10^4$ with the cylinder diameter as the representative length.

\subsection{Magnetic suspension and balance system}\label{sec:MSBS}
The 0.3-m magnetic suspension and balance system (MSBS) at Tohoku University was used as a support system for the model. It consists of a sensor subsystem that monitors the position and attitude of the model, a coil subsystem that generates the magnetic field necessary for levitating and supporting the model, and a control subsystem that connects these two systems to control the position and attitude of the model. The sensor subsystem consists of five charge-coupled device (CCD) line sensors, nine blue light-emitting-diode (LED) light sources, short-pass optical filters, plano-convex lenses and half mirrors. This structure was also adopted in the previous studies \citep{yokota2021analysis, yokota2022instability, yokota2023effect, tashiro2022slanted}. On the other hand, another configuration can be adopted when the fineness ratio is low \citep{inomata2023model, kuwata2021flow}. The CCD line sensors capture the surface of the model illuminated by the LED light sources, and the images are used to detect the edges of the model and the black band painted on the model. After that, the position and attitude of the model are calculated based on the positions of the marker and edges in the image. The relationship between them is obtained in advance by sensor calibration. Subsequently, the calculated position and attitude of the model are used in feedback control, which is designed so as to remove any steady state error. This control subsystem determines the electric current values for each coil to keep the position and attitude of the model close to the set target values. The coil subsystem consists of eight iron-core coils and two air-core coils, which are arranged around the test section. The model is levitated and supported by the interaction between the magnetic field produced by this coil subsystem and the permanent magnets inside the model. The control frequency is 1250 Hz, and the sequence of operations mentioned above is performed within the cycle. The real frequencies of fluctuations by the recirculation bubble pumping and the large-scale vortex shedding, which is predicted from the results of previous studies and the freestream velocity in the present study, are 4.2 and 27.3 Hz, respectively. The control frequency of the MSBS is sufficiently high compared to these frequencies.

The 0.3-m MSBS is capable of position and attitude control in up to six degrees of freedom. Since a freestream-aligned circular cylinder is an axisymmetric body, the wind tunnel tests were conducted to control the model's position and attitude for five degrees of freedom, excluding the roll direction. The model was supported in the centre of the test section and was not rotated significantly in the roll direction during wind tunnel tests. The model position and attitude were acquired by the MSBS, and the model oscillations were evaluated from the data for 7.2~s, corresponding to 9000 sampling points. The root-mean-square (RMS) values of fluctuations in the $x$, $y$, $z$, pitch and yaw directions under the wind-on condition of a run were 4.91~{\textmu}m, 3.13~{\textmu}m, 5.53~{\textmu}m, 0.017~deg and 0.010~deg respectively, which are very small and the effect can be considered negligible.

\subsection{Stereo PIV}\label{sec:PIV}
\begin{figure}
    \centering
    \includegraphics[width=13cm]{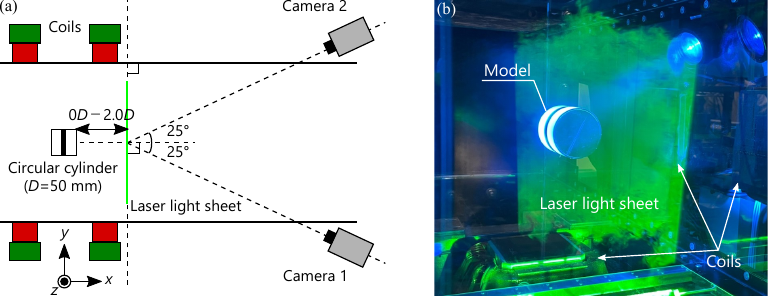}
    \caption{\label{fig:optical} (a) Optical system and configuration of stereo PIV measurements from top. (b) The levitated model with $L/D=1.0$ during measurements, which is viewed from the downstream side of the MSBS.}
\end{figure}

Velocity fields on the $yz$ plane were measured by stereo PIV measurements and the characteristic flow structures were investigated. Figures~\ref{fig:optical} (a, b) show a schematic of the optical system setup for PIV measurements and the levitated model during measurements. The optical system consists of two high-speed cameras (SA-X2, Photron), single focal length lenses (Micro-Nikkor 105 mm f/2.8), bandpass optical filters (527$\pm$10 nm, Edmund Optics), one-axis scheimpflug mounts (Dantec Dynamics), an Nd: YLF laser (LDY-303PIV, Litron), and mirrors (custom-made, SIGMAKOKI). The pixel size, number of pixels and bit depth of the high-speed cameras are 20 {\textmu}m, 1024 x 1024 and 12 bit, respectively. The tracer particles were made of dioctyl sebacate microparticulated by Ruskin nozzles. Note that previous studies have reported that the tracer particles have sufficient followability to the flow \citep{yokota2022instability}.

\begin{figure}
    \centering
    \includegraphics[width=13cm]{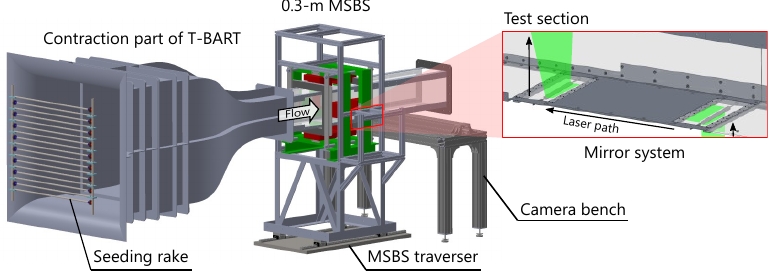}
    \caption{\label{fig:system} Whole view of the developed system for stereo PIV meassurements with the 0.3-m MSBS}
\end{figure}

Optical access for PIV measurements is limited because the model is surrounded by the coils, LED light sources and line sensors. The PIV measurement in the plane parallel to the airflow, which has been employed in previous studies of our group, was relatively easy because the coils in the freestream direction are air-core coils, and the laser sheet could be illuminated from downstream of the test section and a camera could take particle images from between the coils. However, PIV measurements in the $yz$ plane are not possible with the conventional experimental equipment. Therefore, a stereo PIV measurement system, as shown in Fig.~\ref{fig:system}, was developed and introduced that avoids the optical access limitations caused by the MSBS. The newly introduced systems are a mirror system for reflecting the laser light sheet, a traverser to place the MSBS in the desired position and a seeding rake to introduce the particles uniformly. The mirror system consists of two slender mirrors built into the bottom of the test section as shown in Fig.~\ref{fig:system}, each of which can be manually adjusted in angle. When the laser light sheet is illuminated, the laser head is placed at the bottom of the test section facing upwards and is reflected twice by the mirrors to set the measurement plane on the $yz$ plane. The measurement plane is fixed to the test section because of the complex optical setup. Therefore, an MSBS traverser was introduced and the model position relative to the measurement laser plane was changed. The MSBS traverser consists of a horizontal plate and two linear guides, on which the MSBS can be placed for smooth movement in the direction of the wind tunnel axis. The seeding rake consists of two main pipes and 17 sub-pipes and is installed in front of the wind tunnel inlet. Compressed air containing particles produced by the Ruskin nozzles passes through the main pipe and exits through holes in the sub pipe in a spray pattern, thereby introducing particles uniformly throughout the test section. Although the number of particles in the test section increased compared with those in the previous version, the uniform distribution of particles meant that the influence of the scattered light from the LED light source for MSBS was small, and no clear problems occurred with the measurement of the model position and attitude by the sensor system.

The measurement plane was located at 1.0$D$, 1.4$D$ and 2.0$D$ from the cylinder base for all $L/D$ cases. The cameras, as shown in Fig.~\ref{fig:optical}, were positioned at an angle of 25~deg to the wind tunnel axis, looking into the inside of the air-core coil of the MSBS from downstream of the measurement plane. In the camera calibration before measurement, a calibration plate with white dots on a black background was fixed to an automated stage (OSMS20-35(X)-4M4, custom-made, SIGMAKOKI) and calibration images were acquired in 0.1~mm steps over a range of -2~mm $\le x \le$ 2~mm centred on the laser light sheet. The camera calibration was performed only once, as the laser sheet position and camera were fixed. The measurement frequency of the velocity field was set to 400~Hz for all measurement planes. Particle images were acquired five times at 400~Hz for 4,000~pairs (10~s) and once for 10,000~pairs (25~s). However, 0.3~s just after the start of the measurement was not used in the analysis because the laser sheet was not sufficiently bright.

\section{Analysis}\label{sec:Analysis}
\subsection{Velocity field estimation}\label{sec:EstimateVel}
The instantaneous velocity field was calculated by the conventional spatial correlation (CSC) method using analysis software (Dynamic Studio 6.11 and 7.5, Dantec Dynamics). This method was developed by \citet{willert1991digital}. First, a background image was calculated from particle images obtained by each camera. The particle images show the model in the background, and if the cross-correlation of the luminance values between the paired images is taken as is, error vectors will occur near the model. The background image was subtracted and this problem was avoided by eliminating the reflected light on the model. A recursive correlation method was applied to the particle images after background subtraction to calculate the displacement of the group of particles on the image of each camera. The initial size of the correlation window was $32\times32$~px$^2$ and the final size was $8\times8$~px$^2$. A two-dimensional three-component velocity field was then obtained from the displacements of the group of particles in each camera and the results of the calibration. Error vector processing using velocity vectors at eight points around the inspection vector was applied to this velocity field. The velocity field obtained through the above process was used as the instantaneous velocity field for subsequent analysis. The distance between adjacent vectors of the obtained velocity field is 1.27~mm.

\subsection{Transformation of coordinate system}\label{sec:TransCoordinate}
The discussion in the present study will be carried out in Cartesian and cylindrical coordinate systems, as described in \S~\ref{sec:Model}. The Cartesian coordinate system was defined in the software for velocity field estimation, but as the coordinate system defined by the calibration plate and the coordinate system based on the levitated cylinder do not exactly match, a Cartesian coordinate system based on the cylinder was obtained by applying the origin correction based on the time-averaged streamwise velocity. The transformation of the coordinates to a cylindrical coordinate system was done by post-processing in MATLAB. The distance between adjacent vectors in the $r$ direction $\Delta r$ was set to the same value as in the Cartesian coordinate system, with a grid of $N_r = 47$ points in the $r$ direction and $N_\theta = 128$ points in the azimuthal direction. Here, no grid point was placed because the origin is a singularity. The velocity was also transformed from the $y$ and $z$ components to the $r$ and $\theta$ components using the following equation in accordance with the coordinate transformation.
\begin{eqnarray}
    \begin{bmatrix} 
    u_x \\ u_r \\ u_\theta 
    \end{bmatrix}
    &=&
    \begin{bmatrix} 
    1 & 0 & 0 \\
    0 & \cos{\theta} &  \sin{\theta} \\
    0 & -\sin{\theta} & \cos{\theta} 
    \end{bmatrix}
    \begin{bmatrix} 
    u_x \\ u_y \\ u_z
    \end{bmatrix}.
    \label{eq:VelTrans}
\end{eqnarray}
The velocities in the $y$ and $z$ directions at the set cylindrical coordinate grid points were obtained by interpolation from the velocity data in the Cartesian coordinate system and then transformed into the $r$ and $\theta$ components using Eq.~\ref{eq:VelTrans}.


\subsection{Modal decomposition}\label{sec:ModeDecompose}
The wake of the freestream-aligned circular cylinder is a highly complicated flow field due to the convection of KH vortices in the separated shear layer and the relatively large vortex structures including them \citep{yokota2021analysis, yokota2022instability}. It is effective to identify the flow modes which correspond to the fluctuations due to the recirculating bubble pumping and the large-scale vortex shedding from them. Modal decomposition, which combines azimuthal Fourier decomposition and proper orthogonal decomposition (POD) \citep{johansson2006far2, nidhan2020spectral}, is applied to velocity data in a cylindrical coordinate system in the present study to discuss the unexplained phenomena described in the introduction.

The spatial modes and coefficients of the POD at each azimuthal wavenumber were obtained for the acquired velocity data by the following analysis. \reviewerA{The results of time-averaged velocity field and turbulent kinetic energy distribution behind a freestream-aligned circular cylinder in the previous studies have shown homogeneous distributions in the azimuthal direction. In addition, \citet{johansson2006far2} and \citet{nidhan2020spectral} mentioned that the turbulent wake of a disk is homogeneous and periodic in the azimuthal direction. Fourier modes in a homogeneous direction are known to represent POD modes \citep{berkooz1993proper, freund2009turbulence}}. Therefore, the azimuthal Fourier decomposition with a fast Fourier transform is applied to the velocity fields:
\begin{eqnarray}
    {\bf{u}}'(x; r, \theta, t) = \sum_m \Tilde{\bf{u}}_m(x; r, t)e^{im\theta}.
    \label{eq:amd}
\end{eqnarray}
The obtained Fourier coefficients are used to form the following matrix $\Tilde{\bf{U}}_m$ as follows:
\begin{eqnarray}
    \Tilde{\bf{U}}_m = \left[\Tilde{\bf{u}}_m^{(1)}, \Tilde{\bf{u}}_m^{(2)}, \Tilde{\bf{u}}_m^{(3)}, \cdots, \Tilde{\bf{u}}_m^{(N)}\right],
    \label{eq:array}
\end{eqnarray}
where $N$ is the number of instantaneous velocity fields used for POD, which is equal to the number of samples taken in one run of PIV measurements. Since the three velocity components are stacked in the row direction, the size of $\Tilde{\bf{U}}_m$ is $3N_r \times N$. The spatial modes, square roots of eigenvalues and mode coefficients of the POD were obtained using singular value decomposition:
\begin{eqnarray}
    {\bf{W}}\Tilde{\bf{U}}_m = {\bf{U}}_m{\bf{S}}_m{\bf{V}}_m^* = {\bf{U}}_m{\bf{Z}}_m,
    \label{eq:svd}
\end{eqnarray}
where ${\bf{U}}_m$, ${\bf{S}}_m$, and ${\bf{V}}_m$ are matrices including the spatial modes, square roots of eigenvalues and mode coefficients, respectively. ${\bf{Z}}_m$ is used for frequency analyses, conditional sampling and causality analyses in the present study. The asterisk represents complex conjugate transpose. In addition, $\bf{W}$ is the diagonal matrix for weighting with the size of $3N_r \times 3N_r$. Elements of $\bf{W}$ are calculated by the following equation:
\begin{eqnarray}
    W(r) = \sqrt{A(r)} = 
    \sqrt{\frac{\pi\{(r+\Delta r/2)^2-(r-\Delta r/2)^2\}}{2\pi/\Delta\theta}},
    \label{eq:W}
\end{eqnarray}
$A(r)$ is the area that the point at each $r$-position is responsible for, and is used to weight for each of the obtained Fourier coefficients of the three velocity components.


\section{Results and discussions}\label{sec:Result}
\subsection{Flow properties}\label{sec:FlowProperty}
\begin{figure}
    \centering
    \includegraphics[width=13cm]{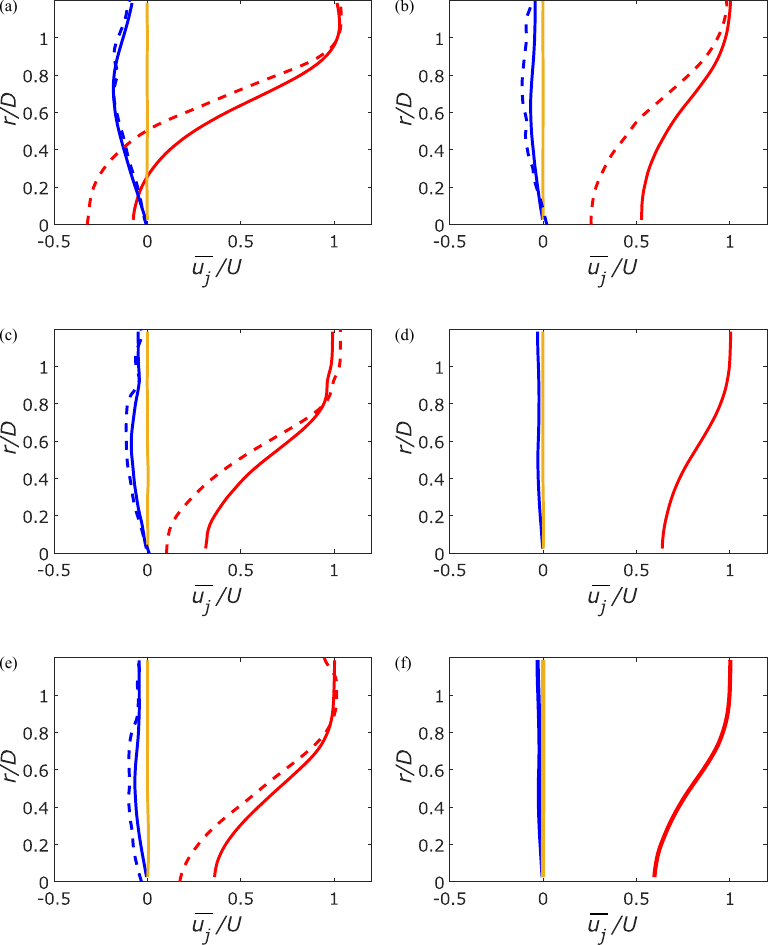}
    \caption{\label{fig:u_mean} The time-averaged velocity profiles for each component (red: $x$, blue: $r$, yellow: $\theta$)} in the case of (a, b) $L/D=1.0$, (c, d) 1.5 and (e, f) 2.0 at (a, c, e) $x/D=1.0$ and (b, d, f) 2.0. The solid lines and the dotted lines represent the results in the present study and the previous study \citep{yokota2021analysis}, respectively.
\end{figure}

Figures \ref{fig:u_mean} show the temporal and azimuthal averaged velocity profiles at $x/D$ = 1.0 and 2.0 for the wake of the cylinder with each $L/D$. The results for freestream and radial velocities are compared with the PIV results of Yokota et al. (two dimensional two component), except for $x/D$ = 2.0 for $L/D$ = 1.5 and 2.0, respectively. They compared the length of the recirculation region behind the cylinder with that behind the disc obtained in previous studies and found good agreement, therefore the data obtained by them are reliable. The figures illustrate that the velocity in the freestream direction in all cases deviates significantly in the positive direction compared to the previous study, with a maximum difference of approximately 30\% of the freestream velocity, even though the trend is in agreement with that of the previous study. The error is considered to be due to the small size of the particle on the image. The effect of peak locking is more apparent when the particle image is small, resulting in bias errors. A camera arrangement such as in the present study, where the camera captures the base of the cylinder, increases the distance between the measurement plane and the camera, leading to a reduction in the spatial resolution. The measurements were carried out with the highest possible lens f-value for preventing peak locking, but the f-number was set to 8 considering the brightness of the particle image and the effect of reflected light on the model. Particularly, the low spatial resolution in the freestream direction in the stereo PIV measurements of the present study is considered to be responsible for the large errors.
The two components corresponding to the in-plane components are in good agreement with the results of the previous study for the $r$ component, while the $\theta$ component is almost zero regardless of the $r$-position, indicating the axisymmetry of the wake of the cylinder. The velocity profiles of the previous study show an unnatural high-wavenumber oscillation in some places due to the time during the measurement when particles could not be introduced into the freestream region and the scratched lines on the acrylic wall used in the test section.

\begin{figure}
    \centering
    \includegraphics[width=12cm]{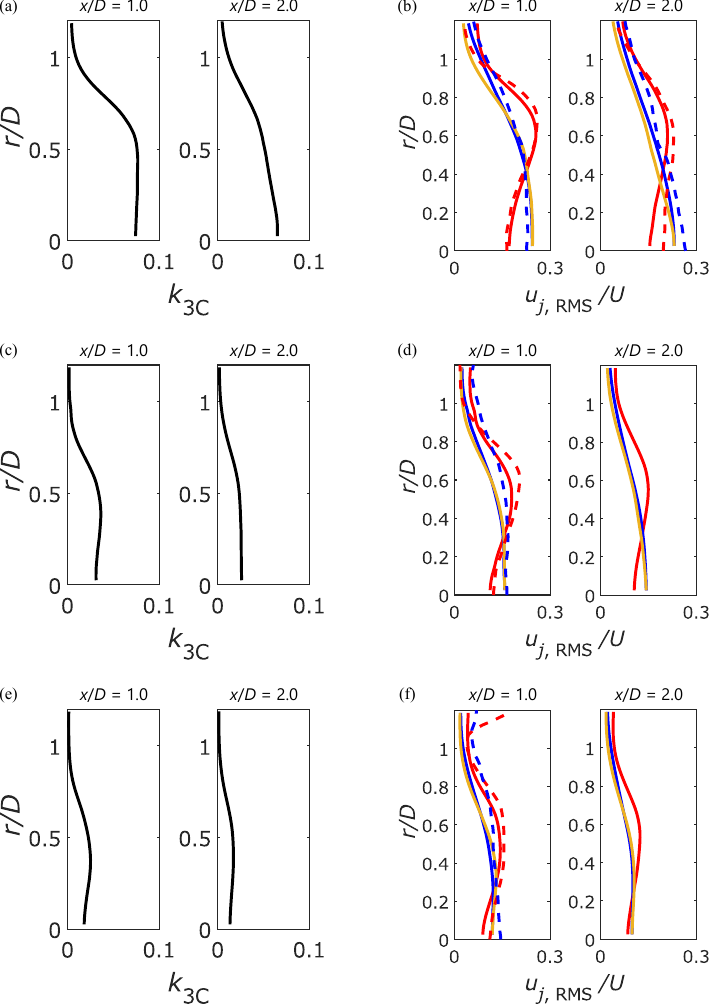}
    \caption{\label{fig:tsq} The profiles of (a, c, e) the turbulent kinetic energy $k_{\rm 3C}$ and (b, d, f) the RMS of velocity fluctuations $u_{j, \rm RMS}$ for each component (red: $x$, blue: $r$, yellow: $\theta$)} in the case of (a, b) $L/D=1.0$, (c, d) 1.5 and (e, f) 2.0. The solid lines and the dotted lines represent the results in the present study and the previous study \citep{yokota2021analysis}, respectively.
\end{figure}

Next, the profiles of the turbulence statistics are discussed. Figure.~\ref{fig:tsq} shows the profile of turbulent kinetic energy $k_{\mathrm{3C}}$ calculated from the three components measured in the present study and the RMS values of fluctuations of each velocity component are shown. The calculations are as follows:
\begin{eqnarray}
    \label{eq:TKE3}
    k_{\mathrm{3C}} &=& \frac{1}{2}\sum_j(u_{j, {\rm{RMS}}}/U)^2 \;(j = x, r, \theta),\\
    \label{eq:u_rms}
    u_{j, {\rm{RMS}}} &=& \sqrt{\frac{1}{N}{\sum_{n=1}^{N}}{u'_{j, n}}^2}.
\end{eqnarray}
All statistics for the data obtained in the present study are averaged in the azimuthal direction. The RMS values are compared with the results of the previous study in the same way as with the case of time-averaged velocity profiles.

Figures~\ref{fig:tsq} (a, c, e) show that $k_{\mathrm{3C}}$ is large at $r/D<0.5$, which is behind the cylinder and decreases rapidly in the freestream region. This tendency is more noticeable at $x/D=1.0$, which is close to the cylinder. In addition, for a given $L/D$, $k_{\mathrm{3C}}$ decreases over the entire $r$-position downstream, and the profile also becomes linear from a curve with a large change from the freestream region to the wake. This is considered to indicate that small vortices or large vortex structures in the turbulence of the wake of the cylinder are convected and weakened by viscous dissipation. Furthermore, a comparison of Figs.~\ref{fig:tsq} (a, c, e) illustrates that $k_{\mathrm{3C}}$ decreases as $L/D$ increases. The reason for this $L/D$ dependence is considered to be related to the reattachment of flow separated at the leading edge to the curved surface of the cylinder. A nonreattaching flow with no flow reattachment occurs at $L/D=1.0$, while a reattaching flow is formed at $L/D=1.5$ and 2.0. This flow classification was made by \cite{yokota2021analysis} based on the time-averaged velocity field, and they reported that the presence or absence of flow reattachment switches at $L/D=1.5$. Note that there is intermittency in flow reattachment for $L/D=1.5$. \cite{higuchi2008sting} reported that the vortex structure formed in the separated shear layer in the case of the reattaching flow is smaller than those in the nonreattaching flow and that the instantaneous velocity also decreases as it approaches the trailing edge. \cite{yokota2021analysis} showed the change by $L/D$ in power spectral densities of velocity fluctuations that are consistent with the report of \cite{higuchi2008sting} and discussed that small vortices formed in the shear layer produce low velocity fluctuations behind the cylinder. The process by which small eddies are formed can be considered as follows. \cite{yokota2022instability} reported that KH vortices in the separated shear layer grow,  merge, and form a relatively large vortex structure, and $Q$-criterion distributions and PSD showed that this vortex structure is even formed before reattachment in the case of $L/D=1.5$ and 2.0. Reattachment of the flow could result in this vortex impinging on the curved surface, collapsing the large-scale vortex structure and breaking it up into smaller vortices. The kinetic energy of the flow may also be reduced by viscous drag due to the formation of a turbulent boundary layer on the surface downstream of the reattachment point. Hence, the velocity fluctuations are considered to be smaller in the case of the reattaching flow than in the case of the nonreattaching flow because the smaller vortices with reduced kinetic energy are shed from the trailing edge of the cylinder. In short, $k_{\mathrm{3C}}$ is considered to be smaller in the order of $L/D=1.0$, where a steady large-scale structure is observed, $L/D=1.5$, where an intermittent large-scale structure is observed, and $L/D = 2.0$, where the vortex structure is small.

The profiles of RMS values of the velocity fluctuations in each direction are shown in Figs.~\ref{fig:tsq} (b, d, f). The results of \cite{yokota2021analysis} are plotted together in these figures for comparison. However, as data for $x/D=2.0$ could not be obtained for the cases $L/D=1.5$ and 2.0 in the previous study, comparisons were not made at that position. As shown in Figs.~\ref{fig:tsq} (d, f), for $L/D=1.5$ and 2.0, the magnitude of the radial velocity fluctuations differs from the previous study for $r/D > 0.6$, while the profiles are consistent for the other positions and cases. Unnatural changes in the $r$ direction are also observed, which could have been caused by problems in PIV measurements due to the particle introduction and the scratched lines on the acrylic wall in the previous study, as described above. The streamwise velocity fluctuations are larger at locations where the change in the time-averaged streamwise velocity in the $r$ direction is greater, i.e. at the shear layer location, and are particularly large for $L/D=1.0$, which is in the case of the nonreattaching flow. On the other hand, the radial and circumferential velocity fluctuations are large at the wake centre and decrease towards the $r$ direction positive. These velocity fluctuations in each direction show a profile similar to the normalised turbulent stresses in the disc wake reported by \cite{nidhan2020spectral}. Here, their paper shows results for $20<x/D<120$. The change in profile in the $r$ direction becomes more gradual downstream of any $L/D$, resulting in a linear turbulent kinetic energy distribution downstream, as shown in Figs.~\ref{fig:tsq} (a, c, e).

\subsection{Characteristic fluctuations}\label{sec:Fluctuation}
\subsubsection{Eigenspectra}\label{sec:EigenSpectra}
\begin{figure}
    \centering
    \includegraphics[width=13cm]{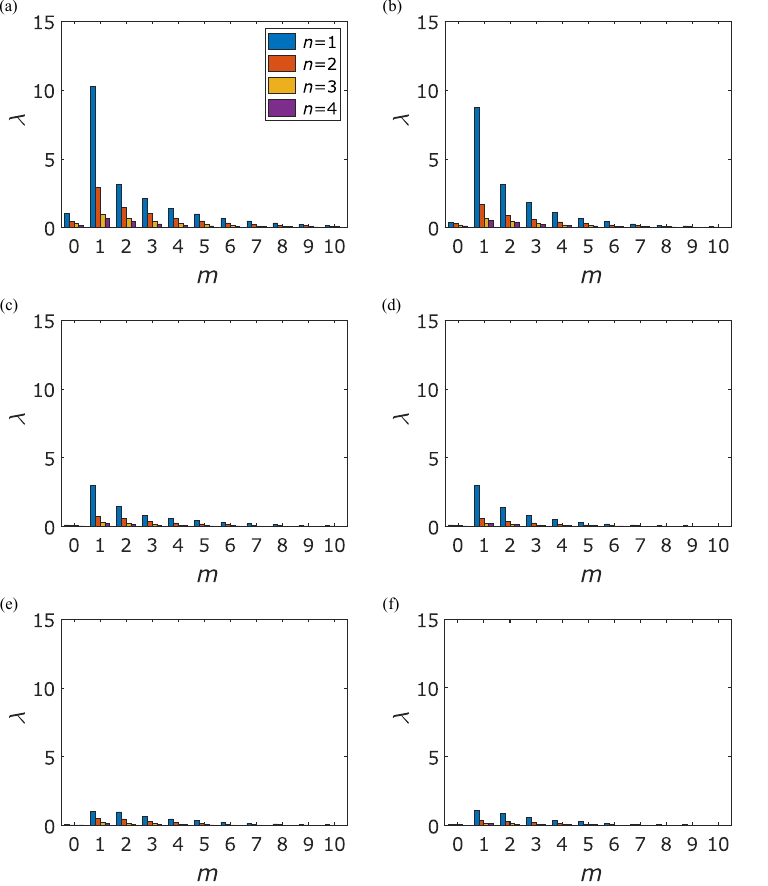}
    \caption{\label{fig:eigenSpectra} The eigenspectra in the case of (a, b) $L/D=1.0$, (c, d) 1.5 and (e, f) 2.0 at (a, c, e) $x/D=1.0$ and (b,d,f) 2.0.}
\end{figure}

Figure~\ref{fig:eigenSpectra} shows the eigenspectra obtained for each $L/D$ and $x/D=1.0$ and 2.0 by applying the modal decomposition described in \S~\ref{sec:ModeDecompose} to the velocity fluctuation field. The range is for the azimuthal mode $m=0\text{--}10$ and leading POD mode $n=1\text{--}4$. At first, eigenvalues become smaller as $L/D$ increases. This is a similar $L/D$-dependent trend to that of $k_{\mathrm{3C}}$ shown in the figure, corresponding to the large velocity fluctuations in the nonreattaching flow and the smaller fluctuations in the reattaching flow. Regardless of $L/D$ and $x$-position, the dominant mode is the $m=1$ mode, which is larger than the other azimuthal modes, especially for $L/D=1.0$ and 1.5. On the other hand, for $L/D=2.0$, the magnitudes of the eigenvalues of $m=1$ and the second contributor, $m=2$, are comparable. The order of contribution of azimuthal modes other than $m=1$ depends on $L/D$ or $x/D$. Comparing the magnitude of the eigenvalues of the $n=1$ mode for each azimuthal mode, the order of contribution for higher azimuthal wavenumbers than the $m=1$ mode decreases with increasing wavenumbers while the order of contribution of the axisymmetric mode $m=0$ varies with $L/D$ or $x/D$. The contribution order of the axisymmetric mode for $L/D=1.0$ is fifth after $m=4$ at $x/D=1.0$, sixth after $m=5$ at $x/D=1.4$ (not shown) and seventh after $m=6$ at $x/D=2.0$.

Since the present study focuses on the large-scale wake structure in the nonreattaching flow, the subsequent discussion will be mainly on the wake behind the cylinder with $L/D=1.0$. \cite{nidhan2020spectral} applied spectral POD to velocity fluctuation fields in the wake of a disc which forms a nonreattaching flow, and showed eigenspectra integrated in the frequency direction. Their results show that the m=0 mode contributes third after the $m=2$ mode at $x/D=0.1$ and fourth after the $m=3$ mode at $x/D\ge1.0$. Here, the results are compared for a disc and a cylinder with $L/D=1.0$, which occur the same flow structure of the nonreattaching flow. \cite{kuwata2021flow} showed that the distance from the leading edge of the cylinder to the downstream end of the recirculation region varies almost linearly with $L/D$ from the results of \cite{fail1957low} and \cite{yokota2021analysis} and them. Although the distance between $L/D=0$ and 1.0 differs by $0.34D$, this difference is not considered to be significant, and the result of \cite{nidhan2020spectral} at $x/D=2.0$ are compared with that of $L/D=1.0$ at $x/D=1.0$, with the same distance from the leading edge. This position is inside the recirculation region in both cases. As mentioned above, the order of contribution of the azimuthal modes for a disc is $1\ge2\ge3\ge0\ge\cdots$, whereas for a cylinder the order is $1\ge2\ge3\ge4\ge0\ge\cdots$ as shown in Fig~\ref{fig:eigenSpectra} (a). This suggests that even if the flow structure is similar, the order of contribution of the azimuthal modes changes with $L/D$.

Next, we consider a large-scale wake structure that corresponds to each mode in the case of $L/D=1.0$. \cite{yokota2021analysis} reported that large-scale vortex shedding appears in the wake at $L/D\le1.5$, which causes velocity fluctuations that are antisymmetric with respect to the cylinder axis. Since the amplitude of the $m=1$ mode is larger than those of the other azimuthal modes at $L/D=1.0$ and 1.5, as mentioned above, and the $m=1$ mode shows an antisymmetric distribution with respect to the cylinder axis, this mode is considered to include the velocity fluctuations by the large-scale vortex shedding. Furthermore, steady symmetry-broken flow, represented by the $m=1$ mode, has been observed in the wake of a disc and an axisymmetric bluff body, which is also considered to be a factor in the dominance of the $m=1$ mode. The next point to note is that the energy of the $m=0$ mode is large only when $L/D=1.0$. The $m=0$ mode is an axisymmetric velocity fluctuation and the relatively high energy in the recirculation region at $x/D=1.0$ suggests that this mode represents recirculation bubble pumping. Furthermore, The $m=2$ mode with a large contribution is considered to correspond to the double-helix structure reported by \cite{johansson2006far2} and \cite{nidhan2020spectral} for the disc wake or the streak-like fluid structure reported by \cite{nekkanti2023large} and \cite{zhang2023coherent}, but as there are no reports on this structure in previous studies on cylinders, it is discussed together with spatial distributions of modes in \S.~\ref{sec:Mode}.

\subsubsection{Eigenfunctions and mode coefficients}\label{sec:Mode}
\begin{figure}
    \centering
    \includegraphics[width=13cm]{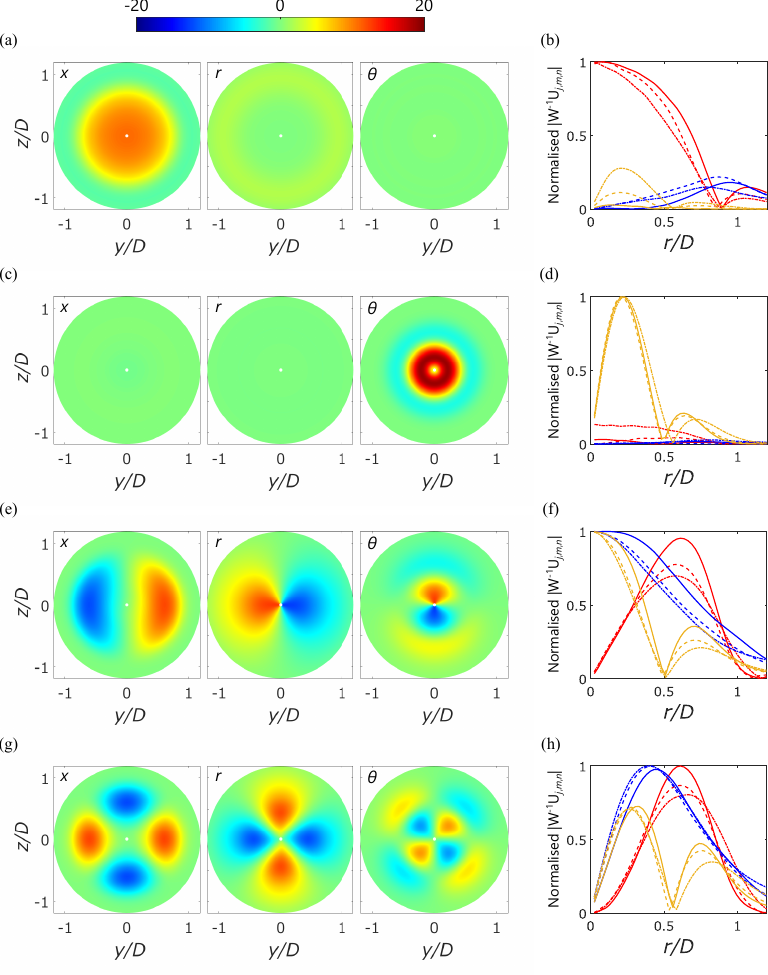}
    \caption{\label{fig:eigenF} The eigenfunctions for mode (a) $(m,n)=(0,1)$, (c) $(m,n)=(0,2)$, (e) $(m,n)=(1,1)$ and (g) $(m,n)=(2,1)$ at $x/D=1.0$ in the case of $L/D=1.0$ and (b, d, f, h) their amplitude normalised by maximum in $|{\bf W}^{-1}{\bf U}_{m, n}|$ at $x/D=1.0$, 1.4 and 2.0, which is expressed by solid lines, dotted lines and single-pointed lines, respectively. The red, blue and yellow lines represent the $x$, $r$ and $\theta$ components, respectively.}
\end{figure}

\begin{figure}
    \centering
    \includegraphics[width=13cm]{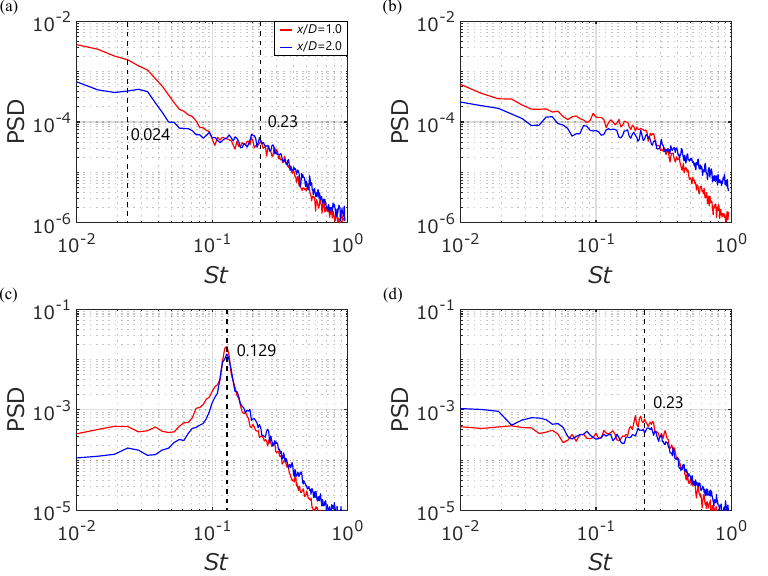}
    \caption{\label{fig:PSD_Z} The power spectral densities of the real part of the mode coefficients for mode (a) $(m,n)=(0,1)$, (b) $(m,n)=(0,2)$, (c) $(m,n)=(1,1)$ and (d) $(m,n)=(2,1)$ at $x/D=1.0$ and 2.0 in the case of $L/D=1.0$.}
\end{figure}

The eigenfunctions of each velocity component at $x/D=1.0$ and 2.0 for mode $(m,n)=(0\text{--}2,1)$ in the case of $L/D=1.0$, which forms the nonreattaching flow focused on in the previous section, are shown in Fig~\ref{fig:eigenF}. In addition, the eigenfunctions for the mode $(m,n)=(0,2)$, which show characteristic spatial patterns of azimuthal shear mode, are also presented. The previous study by \cite{zhang2023coherent} suggested that the planar symmetric vorticity is twisted by azimuthal shear mode in the initial stages of vortex shedding. The contribution of this mode to velocity fluctuations shown in Figs.~\ref{fig:eigenSpectra} (a, b) is small, but it is considered to be related to the large-scale vortex shedding. Therefore, the present study also focuses on it. Velocity fluctuations related to the azimuthal shear mode are also observed for mode $(m,n)=(0,4)$ in the range shown in Fig.~\ref{fig:eigenSpectra}. Figures~\ref{fig:eigenF} (b, d, f, h) show that the eigenfunctions of each mode are almost the same regardless of the $x$-position. Moreover, Fig.~\ref{fig:PSD_Z} shows the power spectral density of the real part of the time-series mode coefficients for these modes at $x/D=1.0$ and 2.0.

Figure~\ref{fig:eigenF} (a) shows that for axisymmetric modes $(m,n)=(0,1)$, the $x$ component is dominant, with large fluctuations in the recirculation region. The $r$ component also fluctuates in phase with the velocity fluctuations in the $x$ direction in the freestream region. On the other hand, the $\theta$ component shows less fluctuation than the other two components, but downstream, fluctuations indicative of flow rotation can be seen in the centre of the wake, which is shown in Fig.~\ref{fig:eigenF} (b). The recirculation bubble pumping is a flow structure that produces axisymmetric velocity fluctuations, and it was reported that large fluctuations appear in the $x$ component for $St<0.05$ \citep{yokota2021analysis}. It is also known from investigations of an axisymmetric bluff body that the recirculation bubble pumping is a local mode appearing in the recirculation region behind the body \citep{zhang2023coherent}.

Fluctuations due to the recirculation bubble pumping are observed at $St\approx0.024$ in the PSD of this mode at $x/D=1.0$ (Fig.~\ref{fig:PSD_Z} (a)). The figure shows that this mode has a difference in the amplitude of fluctuations in the low-frequency region of $St\approx0.024$ inside and outside the recirculation region. In contrast, the fluctuation levels do not change significantly in the other frequency regions. Therefore, the change in fluctuation level in the low-frequency region corresponds to the energy change in the $x$ direction of this mode shown in the eigenspectra (Fig.~\ref{fig:eigenSpectra} (a, b)). The fact that the variation at $St\approx0.024$ is large at $x$-positions corresponding to the inside of the recirculation region suggests that this variation is due to the recirculation bubble pumping \citep{berger1990coherent, yang2015low, yokota2021analysis}, which is a local mode near the recirculation region. The length of the recirculation region is longer than that in the time-averaged field when a negative fluctuation of the $x$ component in the recirculation region occurs. At the same time, the width of the recirculation region also increases, and a negative fluctuation of the $r$ component is considered to be observed in the freestream region, corresponding to flow toward the centre. The PSD of this mode also shows a plateau at $St \approx 0.23$. A plateau at $St\approx0.2$ has been identified in axisymmetric modes in the reports by \cite{nidhan2020spectral} and \cite{nekkanti2023large}, but it is not clear what kind of fluid phenomena they correspond to and its spatial structure. However, as shown by \cite{nidhan2020spectral}, \cite{nekkanti2023large} and the results of the present study, there is a large fluctuation in the frequency around it from near wake to far wake, which is important for the analysis of the wake. The relationship between the fluctuations and them in other modes is discussed in \S~\ref{sec:Relation}.

\begin{figure}
    \centering
    \includegraphics[width=13cm]{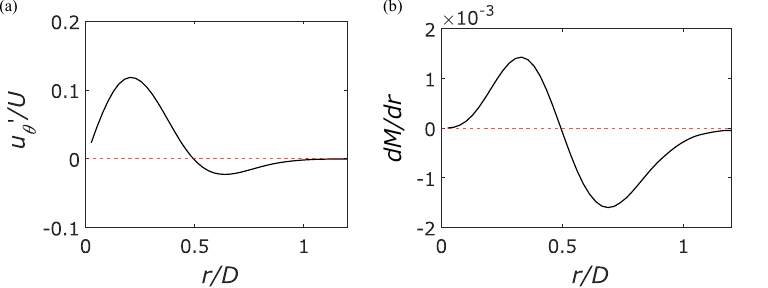}
    \caption{\label{fig:3_moment_02} The $r$ direction profiles of (a) $u_\theta'$ caused by mode $(m,n)=(0,2)$ when $Z_{0,2}=0.01$ at $x/D=1.4$, and (b) its moment $M$ before integration in the $r$ direction.}
\end{figure}

Figure~\ref{fig:eigenF} (c) shows that the $\theta$ component is dominant for the axisymmetric mode $(m,n)=(0,2)$, while the fluctuations of the other two components are very small. The sign of the $\theta$ component is opposite in the centre and outer region, \add{representing a nesting structure of swirl flow in opposite directions. The moments at the cross section are calculated by the following equation \citep{zhang2023coherent}:
\begin{eqnarray}
    \label{eq:moment}
    M(x, t) &=& \int_{r_1}^{r_2}  \int_{0}^{2\pi} \frac{u_\theta(x, r, \theta, t)}{U}\left(\frac{r}{D}\right)^2 \text{d}\theta \text{d}r.
\end{eqnarray}
When $Z_{0,2}=0.01>0$, the moment calculated by the equation is -0.0070 at $x/D=1.4$, which corresponds to the clockwise-direction moment. Figure~\ref{fig:3_moment_02} shows the profiles of $u_\theta'$ and moment in the $r$ direction for this case. The absolute values of the maximum and minimum velocity seem different, and the positive velocity fluctuation appears to be much larger than the negative velocity fluctuation. However, the angular momentum of positive and negative flows, which is obtained by multiplying the moment arm and integrating it in the azimuthal direction, is almost the same as each other. In this case, the angular moment of positive $u_\theta$ fluctuation at $r/D<0.5$ is 0.0136, and the angular moment of negative $u_\theta$ fluctuation at $r/D>0.5$ is -0.0206. This mode was called an azimuthal shear in the previous study as noted below, and the present study illustrates quantitative discussion for the first time.} \cite{yang2014numerical} observed a non-planar-symmetric vorticity distribution called ``Yin-Yang'' pattern in a disc wake, and \cite{zhang2023coherent} suggested that the plane-symmetric vortex may be twisted by azimuthal shear during vortex formation or in the early stages of vortex shedding, forming a ``Yin-Yang'' pattern. The rotational processes were reported to only appear in the recirculation region in the wake of an axisymmetric bluff body, although this cannot be elucidated in the present study as it was only conducted in the region near the recirculation region. The discussion above implicates that this mode of azimuthal shear is considered to be strongly related to vortex shedding and will be further investigated in \S~\ref{sec:VSposition}. The characteristic frequencies have not been reported for velocity fluctuations due to the azimuthal shear mode in previous studies. The PSD of this mode, shown in Fig.~\ref{fig:PSD_Z} (b), decreases in the entire frequency domain as $x/D$ increases, but no peaks are observed, and no characteristic trend is observed, unlike the PSDs of other modes.

Figure~\ref{fig:eigenF} (e) shows that for the antisymmetric mode $(m,n)=(1,1)$, the $u_x$ fluctuation is maximum at the outer edge of the recirculation region. Negative fluctuation regions in the $x$ component correspond to the wake position, with larger fluctuations indicating that the wake position is further away from the central axis of the cylinder. The $u_r$ fluctuations occur at the same position as the $u_x$ fluctuations, but in opposite phase, and this causes the Reynolds stress $-\overline{u_x'u_r'}$. The $\theta$ component has a large fluctuation at a 90-degree deviation in the azimuthal direction from the $x$ and $r$ components. It can be seen that, when considered together with the $r$ and $\theta$ components, the distributions shown in these figures create vortices with opposite sign streamwise vorticity in the $z>0$ and $z<0$ regions. In the region between the two vortices, the flow is from the $u_x$ acceleration region to the deceleration region, and the flow outside each vortex is from the $u_x$ deceleration region to the acceleration region. 

The PSD of this mode shown in Figs.~\ref{fig:PSD_Z} (c) represents that there is a clear peak at $St=0.129$. This fluctuation frequency is in good agreement with the large-scale vortex shedding in the wake of the freestream-aligned circular cylinder reported by previous studies \citep{berger1990coherent, yang2015low, nidhan2020spectral, yokota2021analysis}. Peaks at this frequency appear from near wake to far wake, indicating that large-scale vortex shedding is a global fluctuation \citep{johansson2006far2, nidhan2020spectral, nekkanti2023large}. However, the intensity of the fluctuations weakens as $x/D$ increases, as can be seen from the present study and their results. Similarly, the fluctuations weaken in the low-frequency region of $St\le0.05$ as $x/D$ increases. Low-frequency fluctuations of $m=1$ modes have not been discussed in flow around a freestream-aligned circular cylinder, but the study of the wake of an axisymmetric bluff body \citep{zhang2023coherent} has identified fluctuations with the same spatial structure as very low frequency (VLF) mode, with $St$ on the order of $10^{-2}$. It is known that VLF mode exists as an $m=1$ mode near the object, but switches to an $m=2$ mode around the downstream end of the recirculation region. The relationship between $m=1$ and $m=2$ modes is also discussed in \S~\ref{sec:Relation}.

Figure~\ref{fig:eigenF} (g) shows that mode $(m,n)=(2,1)$, as with the mode $(m,n)=(1,1)$, has large $u_x$ fluctuations at the outer edge of the recirculation region and that the region with large fluctuations also moves in the positive direction of the $r$ axis as $x/D$ increases. The $u_r$ fluctuations are also large in the same position as the $u_x$ fluctuations as well as mode $(m,n)=(1,1)$ and are in opposite phases. Hence, mode $(m,n)=(2,1)$ is also a factor producing Reynolds stress $-\overline{u_x'u_r'}$. The $\theta$ component has a large fluctuation at a position 90 degrees shifted azimuthally from the other two components, and when considered together with the $r$ component, shows that a total of four vortices appear, with the sign of the streamwise vorticity switching alternately in the azimuthal direction. This structure is also seen in the disc wake \citep{nekkanti2023large} and is very similar to that of the present study. There is an acceleration region of $u_x$ at the boundary of vortices with switching positive to negative streamwise vorticity in the azimuthal direction, and a deceleration region of $u_x$ at the boundary of vortices with negative to positive streamwise vorticity. A very similar structure was found by \cite{nekkanti2023large} in their spectral POD (SPOD) analysis of the disc wake, which is a streak structure represented by the first SPOD mode with $m=2$, $St\rightarrow0$. Another similar spatial structure is the double-helix structure reported by \cite{zhang2023coherent}. In their report, the cross-sectional pattern of the streamwise component obtained by dynamic mode decomposition (DMD), which corresponds to the double-helix structure, shows a ``tail'' that looks like the $x$ component pattern in Fig.~\ref{fig:eigenF} (g) twisted in the azimuthal direction. The frequency of velocity fluctuations due to the double-helix structure occurring in the disc wake is $St\approx0.2$ at $x/D=2.0$ \citep{nidhan2020spectral} and $St=0.27$ at $x/D=10$ \citep{nekkanti2023large}. Hence, a large fluctuation in the PSD of this mode is expected at $St\rightarrow0$ and $St=0.2\text{--}0.3$.

The PSDs for this mode shown in Fig.~\ref{fig:PSD_Z} (d) reveal a small broad peak at $St \approx 0.23$. The fluctuations at this frequency become smaller as $x/D$ increases. \cite{zhang2023coherent} present that the double-helix structure is apparent in the wake of an axisymmetric bluff body at almost twice the frequency of large-scale vortex shedding and that the fluctuations decrease downstream, from the spatial distribution of the POD modes. Similarly, in the disc wake, peaks of fluctuation due to the double-helix structure were observed for $x/D=2.0$ \citep{nidhan2020spectral} and 10 \citep{nekkanti2023large}, but not for $x/D\ge20$ \citep{johansson2006far2, nidhan2020spectral}, which suggests that the fluctuations are smaller as $x/D$ increases in the intermediate wake. Therefore, the fluctuations at $St \approx 0.23$ in the cylinder wake are considered to be caused by the double-helix vortex. On the other hand, the fluctuations increase downstream in the low-frequency region of $St\le0.05$, which is opposite to the trend in the fluctuations in the low-frequency region for mode $(m,n)=(1,1)$. Since fluctuations with the same spatial structure as the VLF mode mentioned above make a transition from $m=1$ to $m=2$ around the downstream end of the recirculation region, the reversal of the tendency of the PSDs for $m=1$ and $m=2$ in the low-frequency range to change in the downstream direction is also considered to be indicative of this transition. Another related fluctuation in the low-frequency region of the azimuthal mode $m=2$ is the recirculation bubble pumping. \cite{ohmichi2019numerical} numerically investigated the wake of the re-entry capsule and presented that recirculation bubble pumping is associated with four streaks that appear downstream of the recirculation region by showing corresponding DMD results. In any case, the low-frequency fluctuations of mode $(m,n)=(2,1)$ are considered to indicate four streaks appearing in the wake.

\begin{figure}
    \centering
    \includegraphics[width=13cm]{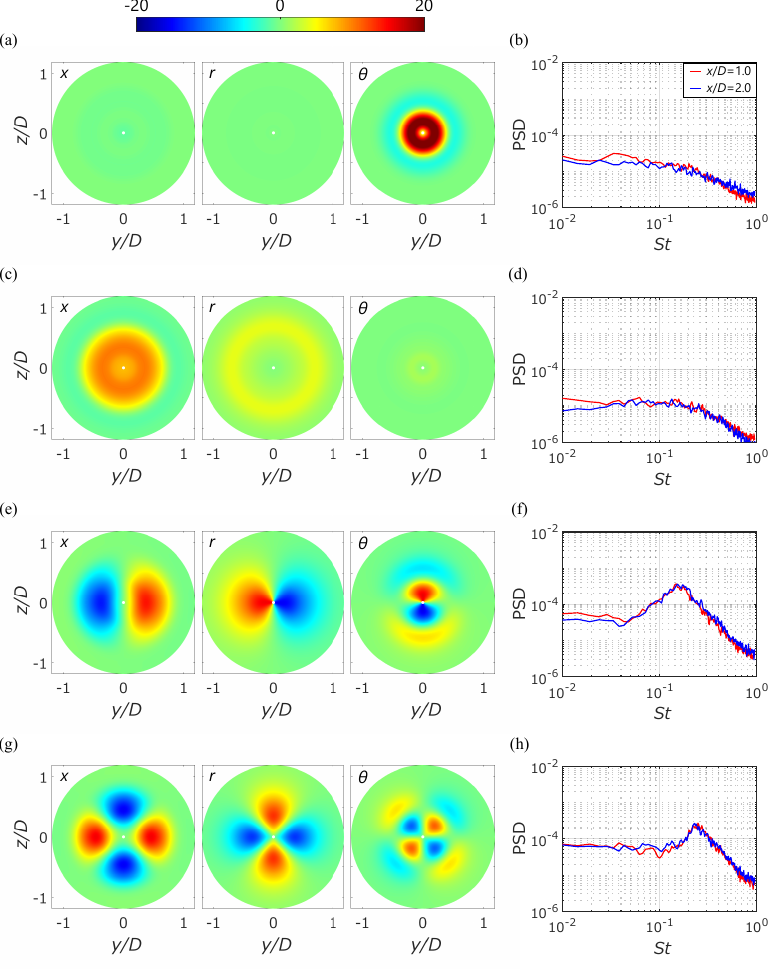}
    \caption{\label{fig:eigenF_lbd200} The eigenfunctions for mode (a) $(m,n)=(0,1)$, (c) $(m,n)=(0,2)$, (e) $(m,n)=(1,1)$ and (g) $(m,n)=(2,1)$ at $x/D=1.0$ in the case of $L/D=2.0$ and (b, d, f, h) PSDs of the real part of the mode coefficients at $x/D=1.0$ and 2.0, which is represented by the blue and red lines, respectively. The red, blue and yellow lines represent the $x$, $r$ and $\theta$ components, respectively.}
\end{figure}

Figure~\ref{fig:eigenF_lbd200} shows the eigenfunctions and the PSDs of the mode coefficients for $L/D=2.0$. Note that both $x/D=1.0$ and 2.0 are the downstream side of the recirculation region because the reattaching flow is formed for $L/D=2.0$ and the recirculation region is short. The eigenfunction for mode $(m,n)=(0,1)$ is similar to those in Fig.~\ref{fig:eigenF} (c) representing the azimuthal shear mode, and the eigenfunction for mode $(m,n)=(0,2)$ is also similar to those in Fig.~\ref{fig:eigenF} (a) representing the bubble pumping, which indicates that the azimuthal shear mode is dominant in axisymmetric fluctuations for the reattaching flow. The mode order is switched compared with that of the nonreattaching flow because the fluctuations due to recirculation bubble pumping are smaller in the case of the reattaching flow, and the energy is relatively low compared with the azimuthal shear mode. The low-frequency fluctuations for mode $(m,n)=(0,2)$, considered to be related to the recirculation bubble pumping, are smaller than the case with $L/D=1.0$ and $x/D=2.0$. The eigenfunctions for mode $(m,n)=(1\text{--}2,1)$ are almost the same pattern as those shown in Figs.~\ref{fig:eigenF} (e, g). Fluctuations due to the large-scale vortex shedding do not appear in a narrow band like in the case of $L/D=1.0$, but in a broadband fluctuation. Fluctuations due to the double-helix structure also appear at $St=0.2\text{--}0.3$. Although not shown in the paper, the results for $L/D=1.5$ are similar to those for $L/D=2.0$. 

\subsection{Change in vortex shedding position}\label{sec:VSposition}
The vortex shedding position has been reported to be related to aerodynamic fluctuations in the lateral direction of a freestream-aligned circular cylinder by \cite{yokota2021analysis} and \cite{shinji2020aerodynamic}. Therefore, the comprehension of the vortex shedding position is important for suppressing vibrations acting on the cylinder. In addition, \cite{yang2014numerical} suggested that fluctuations in the azimuthal position of vortex shedding are associated with the recirculation bubble pumping. The discussion proceeds for the vortex shedding position before investigating this relationship. In the present study, the representative position of the vortex shedding position is expressed by the barycentre of the momentum deficit in the following equation, as used in the previous studies \citep{grandemange2013turbulent, yang2014numerical, gentile2016low, zhang2023coherent}:
\begin{eqnarray}
    \label{eq:ym}
    y_m &=& \frac{\int y(1-u_x/U)dS}{\int (1-u_x/U)dS},\\
    \label{eq:zm}
    z_m &=& \frac{\int z(1-u_x/U)dS}{\int (1-u_x/U)dS},
\end{eqnarray}
where $S$ is the examined area, which in the present study is [-1.2$D$ 1.2$D$] in both $y$ and $z$ directions. Figure~\ref{fig:u_mean} shows that the time-averaged velocity in the freestream direction is positively biased, but that the fluctuation components in the present study are in good agreement with the previous study. Therefore, the qualitative discussion should be reasonable.  

\begin{figure}
    \centering
    \includegraphics[width=13cm]{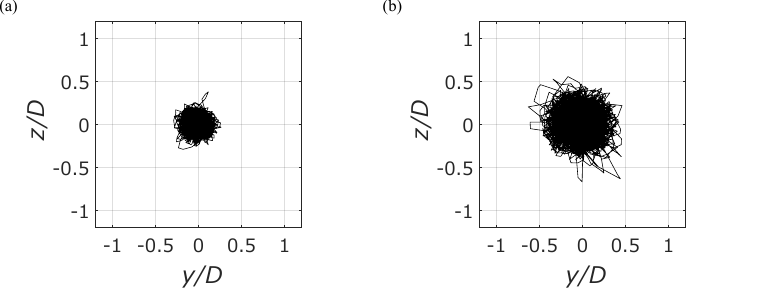}
    \caption{\label{fig:wake_pos} The trajectories of the wake position at (a) $x/D=1.0$ and (b) 2.0 in the case of $L/D=1.0$.}
\end{figure}

\begin{figure}
    \centering
    \includegraphics[width=13cm]{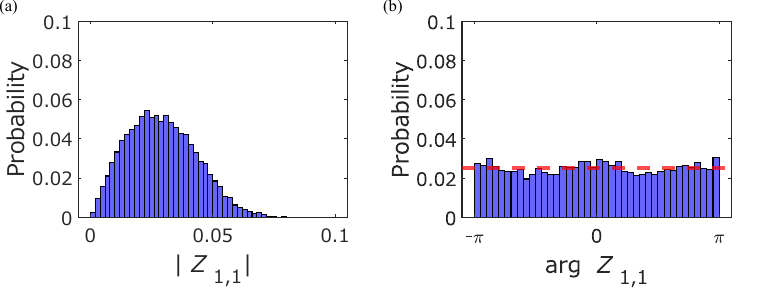}
    \caption{\label{fig:prob_amp_ang} The probability distribution of (a) the amplitude and (b) argument of the mode coefficients for mode $(m,n)=(1,1)$.}
\end{figure}

Figure~\ref{fig:wake_pos} shows that the vortex shedding position fluctuates irregularly around the origin. Here, the trajectory is based on the data for 24.7~s. They also show that the circular area drawn by the trajectory becomes larger as $x/D$ increases, indicating that the wake width increases. The centre of the trajectory is shifted from the origin to the negative direction of the $y$ axis for $x/D=2.0$ (Fig.\ref{fig:wake_pos} (b)), which is due to error vectors at the left end of the examined area. Figure~\ref{fig:prob_amp_ang} shows the probability distributions obtained from the amplitude and argument of mode $(m,n)=(1,1)$ for 24.7~s at $L/D=1.0$ and $x/D=1.4$, which are considered to correspond to the radial and azimuthal positions of the wake position, respectively. Here, the correlation coefficient between the wake radial position $r_m$ and the amplitude of mode $(m,n)=(1,1)$ is 0.72. The distribution of $|Z_{1,1}|$ has a maximum in the bin of 0.022--0.024, as shown in Fig.~\ref{fig:prob_amp_ang} (a), and the maximum not being at zero indicates that the flow field is quasisteadily asymmetric. This steady symmetry breaking has also been observed in the wake of an axisymmetric bluff body \citep{rigas2014low, rigas2015diffusive}. On the other hand, the probability of $\mathrm{arg}~Z_{1,1}$ shown in Fig.~\ref{fig:prob_amp_ang} (b) represents an almost uniform distribution. Here, the red broken line in the figure shows the position where the probability is 0.025. Since the number of bins in the histogram is 40, it should match the red dashed line if the probabilities were uniform, but a slight dispersion is observed. The average and standard deviation of the probability are 0.025 and 0.0026, respectively. The uniformity of $\mathrm{arg}~Z_{1,1}$ means that there is no bias in the azimuthal position of the spatial pattern shown in Fig.~\ref{fig:eigenF} (e) and that the mean velocity field represents an axisymmetric distribution about the cylinder axis. Although results are not shown, the probability distribution of $|Z_{1,1}|$ does not show a maximum at zero in the case of $L/D=1.5$ and 2.0, indicating steady symmetry breaking, and $\mathrm{arg}~Z_{1,1}$ also shows a uniform distribution. Therefore, stationary symmetry breaking appears regardless of shear layer reattachment.

\begin{figure}
    \centering
    \includegraphics[width=13cm]{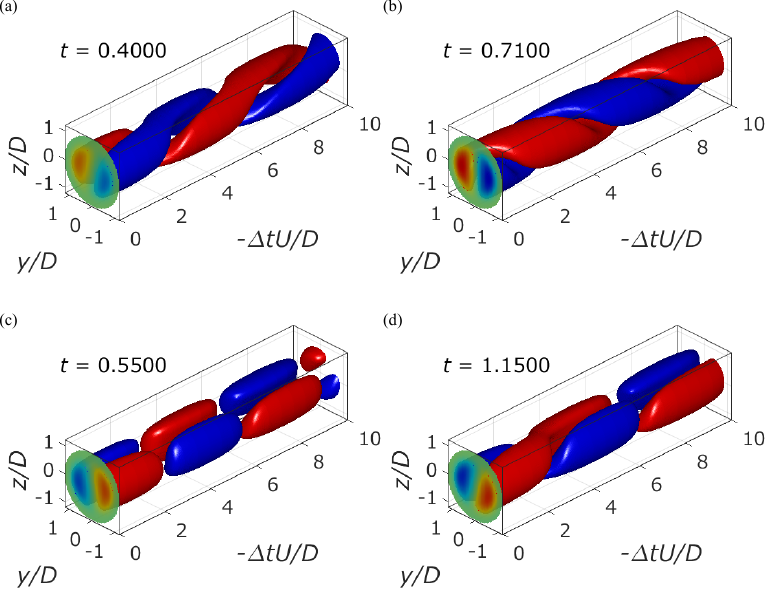}
    \caption{\label{fig:taylor} The snapshots of the pseudo-three-dimensional $u_x'$ map for the mode of $(m,n)=(1,1)$ with the state of (a) anticlockwise circular, (b) clockwise circular, (c) flapping and (d) mixture of circular and flapping at $x/D=1.4$ in the case of $L/D=1.0$.}
\end{figure}

The trajectory of the vortex shedding position represents a loop or a flapping-like reciprocating motion, depending on the time. \cite{yang2014numerical} also showed trajectories of vortex shedding positions under $Re=10^4$, in which closed loops were identified. They suggested that this loop was a helical vortex structure appearing in the wake of the cylinder. However, they also recognised that the shape of the loops varied irregularly with time. If only helical vortices appeared, they would show a circular trajectory as shown by \cite{zhang2023coherent}. Therefore, helical vortices, flapping, and a structure mixing them in the wake can be assumed to appear depending on the time at this Reynolds number. Consequently, the velocity fluctuations due to mode $(m,n)=(1,1)$ at $x/D=1.4$, corresponding to large-scale vortex shedding, are visualised three-dimensionally using Taylor's hypothesis. Bandpass filtering of the mode coefficients with $0.1\le St \le0.2$ was applied, and large-scale vortex shedding was focused. The parameter $k_{3C}^{1/2}$, calculated from the turbulent kinetic energy, is at the maximum of 28.4\% for $L/D=1.0$ and $x/D=1.4$, and although it is not appropriate to apply Taylor's hypothesis as in the paper of \cite{nekkanti2023large}, it was applied because it is useful for understanding changes in the vortex shedding position. Figure~\ref{fig:taylor} shows selected visualisations when helical vortices, flapping or a mixture of both are considered to occur. The left contour plot shows the velocity fluctuations of the $(m,n)=(1,1)$ mode at time $t$, and the isosurface is formed by connecting the points where the fluctuations are $\pm0.05$. Figures~\ref{fig:taylor} (a, b) show anticlockwise and clockwise helices in the view from downstream, respectively, where twisting occurs without significant change in the magnitude of the fluctuations. In other words, the vortex shedding position in the $r$ direction remains the same, but its position changes in the azimuthal direction in a circular pattern. On the other hand, Fig.~\ref{fig:taylor} (c) shows that the positive and negative values switch without changing the azimuthal position, indicating that vortex shedding is performed in a certain plane like flapping motion. Furthermore, Fig.~\ref{fig:taylor} (d) shows a mixture of helical and flapping features and is considered to be an elliptical motion. These four states appear irregularly, as shown in the supplementary material of movie 1. These states are also observed for $L/D=1.5$ and $2.0$ from movies 2 and 3 of the supplementary material, respectively. The isosurfaces in these videos are displayed by connecting the points where the fluctuations are $\pm0.01$.

\begin{figure}
    \centering
    \includegraphics[width=13cm]{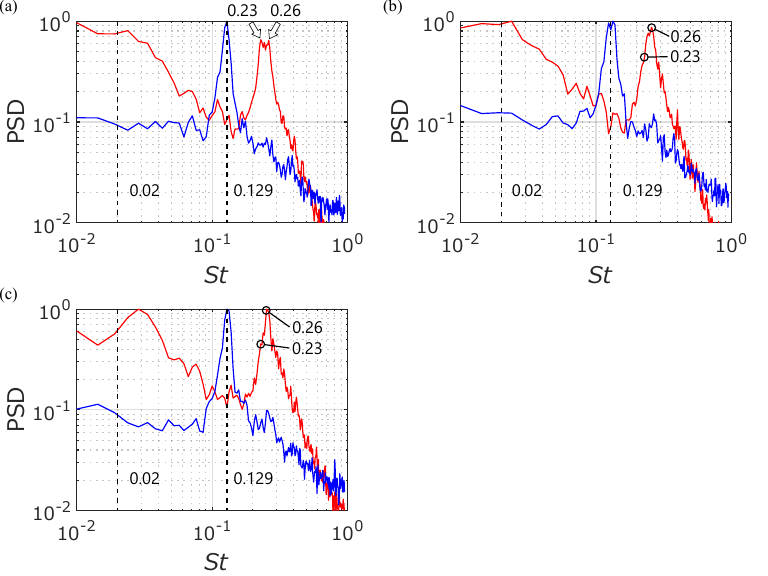}
    \caption{\label{fig:PSD_wake_pos} The normalized power spectral densities of the amplitude (red) and the angular change (blue) of spatial pattern of mode $(m,n)=(1,1)$ at (a) $x/D=1.0$, (b) 1.4 and (c) 2.0 in the case of $L/D=1.0$.}
\end{figure}

Figure~\ref{fig:PSD_wake_pos} shows the PSD calculated from the amplitude and the angular change of the spatial pattern of mode $(m,n)=(1,1)$, which are considered to correspond to the radial and azimuthal fluctuations of the vortex shedding position, respectively. The angular variation of the spatial pattern was calculated from the azimuthal position of the deceleration region of the streamwise velocity corresponding to the vortex shedding position. The data used for the calculation of the PSDs were the same as those used for the plot of the trajectories, while each PSD was normalised by its maximum value. Only azimuthal position fluctuations appear when the trajectory of the vortex shedding position is circular, as shown in Figs.~\ref{fig:taylor} (a, b), and a peak is considered to appear at $St=0.129$ in the PSD of the angular variation. However, the angular variation appears as a fluctuation at $St=0.129$ even in the case of flapping, as shown in Fig.~\ref{fig:taylor} (c). The gradient of the angular change, i.e. the angular velocity, is necessary to classify them. Flapping or elliptical loops across the vicinity of the cylinder axis, as shown in Fig.~\ref{fig:taylor} (c, d), are expected to appear as the twice higher fluctuation frequency due to large-scale vortex shedding. This is because the sign of the velocity switches when the velocity fluctuations due to vortex shedding show a flapping pattern, and $|Z_{1,1}|$ goes from near its maximum value to zero and back to a value near its maximum value again. Thus, when the velocity fluctuations advance by half a period, the fluctuations in $|Z_{1,1}|$ advance by one period, and a peak appears at twice the frequency of the fluctuations due to the large-scale vortex shedding. Accordingly, the fluctuations at $St\approx0.26$ are considered to be due to flapping patterns. The peak at $St\approx0.26$ is observed regardless of $x$-position. A similar spectrum of positional fluctuations has been observed immediately behind an axisymmetric bluff body \citep{zhang2023coherent}. Other large fluctuations were observed, appearing at $St\approx0.02$ and 0.23 of the PSD of the amplitude. The low-frequency fluctuations at $St\approx0.02$ are considered to be related to the recirculation bubble pumping. The vortex shedding position becomes larger or smaller in the $r$ direction as the size of the recirculation region changes in the streamwise direction. At last, the fluctuations at $St=0.23$, which are comparable to the fluctuations at $St\approx0.26$ in the case of $x/D=1.0$, are relatively weak compared to the fluctuations at $St=0.26$ as $x/D$ increases, suggesting that these fluctuations are represented inside the recirculation region. The broadband large fluctuations are observed at $St\approx0.23$ for modes $(m,n)=(0,1)$ and (2,1), as shown in the Fig.~\ref{fig:PSD_Z}, and further discussion of the relationship between the modes at $St=0.23$ will be carried out in the next section.

\begin{figure}
    \centering
    \includegraphics[width=13cm]{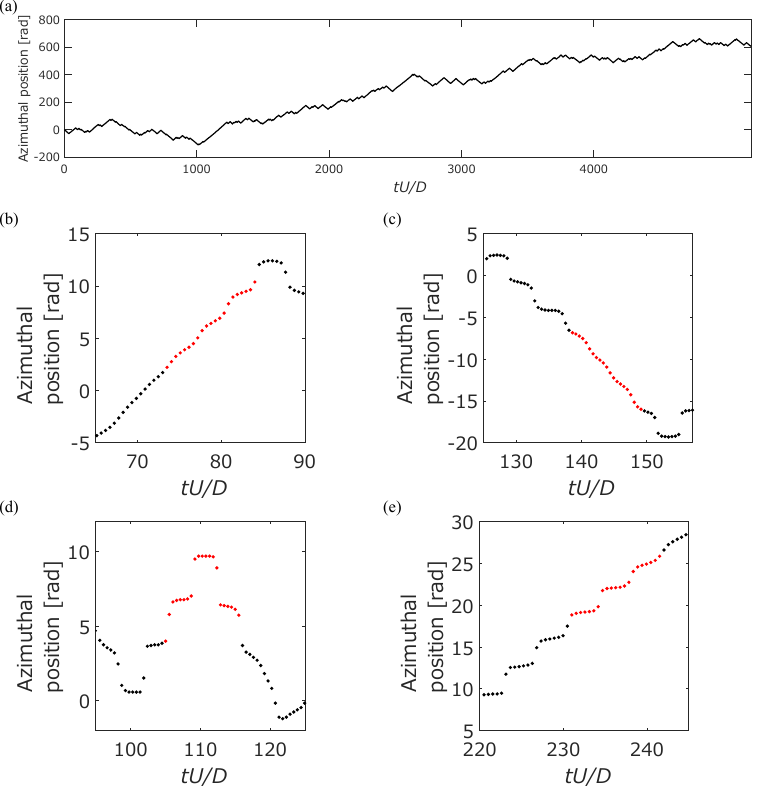}
    \caption{\label{fig:ang} The temporal variation of the vortex shedding position at $x/D=1.4$ for (a) the whole measurement time and (b--e) the time of pseudo-three-dimensional maps shown in Figs.~\ref{fig:taylor} (a--d), respectively. The red dots correspond to the moment of snapshots shown in Fig.~\ref{fig:taylor}.}
\end{figure}

\begin{figure}
    \centering
    \includegraphics[width=13cm]{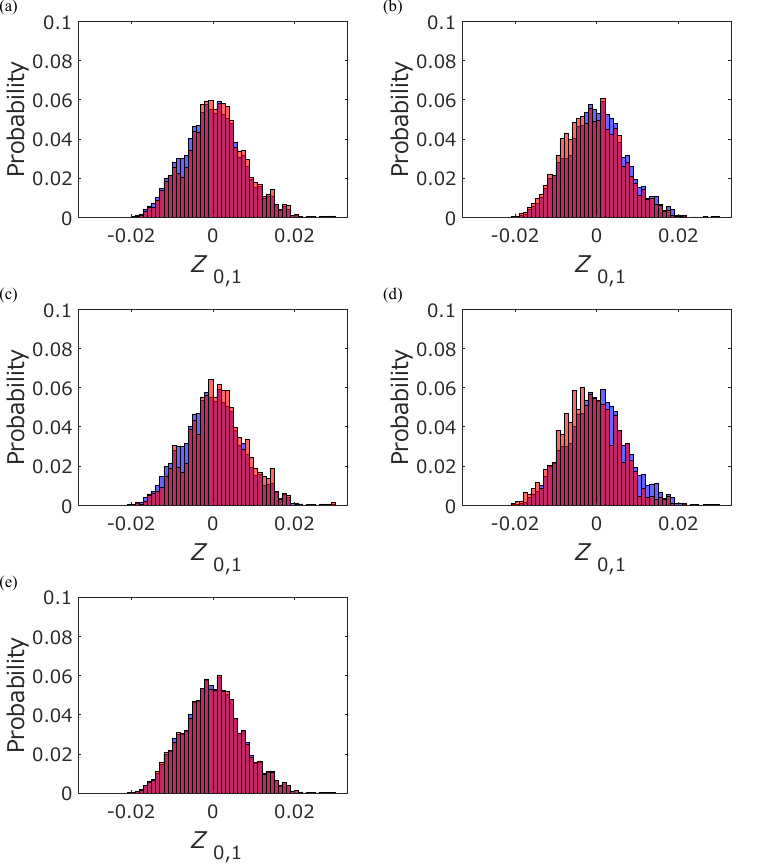}
    \caption{\label{fig:prob_0_1} The probability distribution of $Z_{0,1}$ obtained by sampling under the condition that the state of vortex shedding is (a) anticlockwise, (b) clockwise, (c) anticlockwise circular, (d) clockwise circular and (e) flapping at $x/D=1.4$ in the case of $L/D=1.0$. The red and blue histograms show the results of the conditional sampling and the whole measurement time, respectively.}
\end{figure}

Figure~\ref{fig:ang} shows the temporal variation of the vortex shedding position at $x/D=1.4$, close to the downstream end of the recirculation region. Figure~\ref{fig:ang} (a) plots the azimuthal position over 24.7~s, while Figs.~\ref{fig:ang} (b--e) show the change in the position over time of the pseudo-three-dimensional maps shown in Figs.~\ref{fig:taylor}(a--d), respectively. As in the previous paragraph, bandpass filtered data with $0.1\le St \le0.2$ was used here, and the azimuthal position was calculated. Figure~\ref{fig:ang} (a) shows that the rotational direction of the vortex shedding switches irregularly with a major trend to rotate in the positive direction of the $\theta$ axis at the times shown. Although not shown here for brevity, different runs at the same location had a major trend to rotate in the negative direction of the $\theta$ axis. \cite{yang2015low} suggested that switching the direction of rotation of the vortex shedding position is associated with the recirculation bubble pumping. Therefore, conditional sampling was applied to mode $(m,n)=(0,1)$ and its association was investigated in the present study. Figures~\ref{fig:prob_0_1} (a, b) (red histograms) show the results of sampling $Z_{0,1}$ at the time for positive and negative slopes of the position fluctuations of the vortex shedding. The probability distributions of them over the whole measurement time are shown together in the figure for comparison (blue histogram). The $Z_{0,1}$ probability distributions should be biased if the rotational direction of the vortex shedding position is associated with recirculation bubble pumping, but no such bias is observed in the results. In other words, there is no relationship between the rotational direction of the vortex shedding position and the recirculation bubble pumping.

The present study investigates not only the rotational direction of the vortex shedding position but also whether the state of the recirculation region differs depending on whether the vortex shedding pattern shows a circular or reciprocating pattern, such as flapping. The gradient of the azimuthal position was used as a condition for classifying these patterns. The conditions are as follows:
\begin{eqnarray}
    \label{eq:VSstate}
    &\text{state is}
    \begin{cases}
        \text{anticlockwise} & \text{if}~\frac{\omega_m}{2\pi}\ge0, \\
        \text{clockwise} & \text{if}~\frac{\omega_m}{2\pi}<0, \\
        \text{anticlockwise circular} & \text{if}~0.1\le\frac{\omega_m}{2\pi}\le0.2, \\
        \text{clockwise circular}     & \text{if}~-0.2\le\frac{\omega_m}{2\pi}\le-0.1, \\
        \text{flapping}               & \text{if}~|\frac{\omega_m}{2\pi}|<0.1~\text{or}~|\frac{\omega_m}{2\pi}|>0.2,\\
    \end{cases}\\
    &\omega_m = \frac{d\theta_m}{d(tU/D)},
\end{eqnarray}
where $\theta_m$ is the azimuthal position with the minimum $u_x'$ of mode $(m,n)=(1,1)$. Figures~\ref{fig:ang} (b, c) show that the vortex shedding position varies smoothly if the pattern is circular. On the other hand, Fig.~\ref{fig:ang} (d) illustrates that discontinuous changes and near-zero gradient changes appear when a flapping pattern appears. The present authors compared the pseudo-three-dimensional field shown in Fig.~\ref{fig:taylor} with the positional changes in Fig.\ref{fig:ang} and confirmed that the sample points satisfying the conditions were selected appropriately. The points of the circle pattern and the points of the flapping pattern were confirmed to appear when the two patterns are mixed, as shown in Fig.~\ref{fig:ang} (e). Figures~\ref{fig:prob_0_1} (c--e) show the results of sampling $Z_{0,1}$ when the anticlockwise/clockwise circular pattern and the flapping pattern appear, respectively. In the case of the circular pattern shown in Figs.~\ref{fig:prob_0_1} (c, d), the distribution is slightly different from that of $Z_{0,1}$ for the whole measurement time, but there is no clear difference. No significant changes in the distribution of $Z_{0,1}$ compared to the probability distribution of it over the whole measurement time are also observed in Fig.~\ref{fig:prob_0_1}, suggesting that there is no link between the shedding pattern and the recirculation bubble pumping. 

\begin{figure}
    \centering
    \includegraphics[width=13cm]{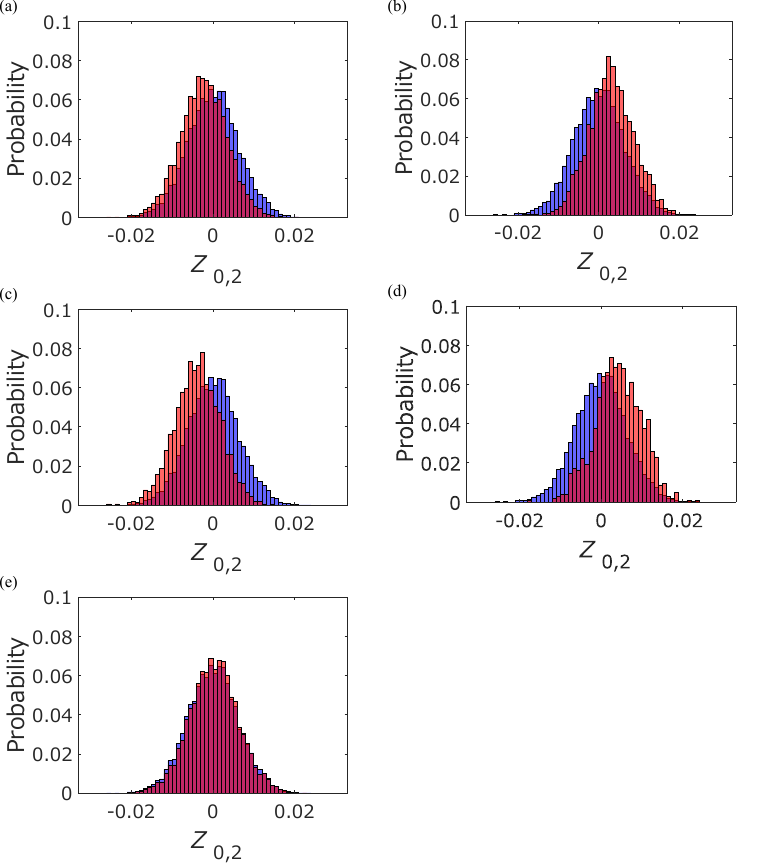}
    \caption{\label{fig:prob_0_2} The probability distribution of $Z_{0,2}$ obtained by sampling under the condition that the state of vortex shedding is (a) anticlockwise, (b) clockwise, (c) anticlockwise circular, (d) clockwise circular and (e) flapping at $x/D=1.4$ in the case of $L/D=1.0$. The red and blue histograms show the results of the conditional sampling and the whole measurement time, respectively.}
\end{figure}

The distribution of azimuthal shear mode is shown at mode $(m,n)=(0,2)$ in the present study, and the previous study \citep{zhang2023coherent} suggests that the mode is highly associated with vortex shedding. \cite{zhang2023coherent} reported that azimuthal shear mode is found within the recirculation region, twisting planar symmetric vortex loops, and exhibiting a vorticity distribution called ``Yin-Yang'' pattern. Here, the pattern is the RSB distribution, which was also found in the disc wake \citep{yang2014numerical}. Therefore, conditional sampling was applied to this mode as well as mode $(m,n)=(0,1)$. The conditions are described in Eq.~\ref{eq:VSstate}. Figures~\ref{fig:prob_0_2} (a--e) show the results of sampling $Z_{0,2}$ for anticlockwise and clockwise changes, anticlockwise and clockwise circular patterns and flapping, respectively, as in Fig.~\ref{fig:prob_0_1}. At first, Figs.~\ref{fig:prob_0_2} (a, b) presents that $Z_{0,2}$ are biased depending on the direction of rotation of the vortex shedding position. It is negatively biased when the vortex shedding is anticlockwise and positively biased when the shedding is clockwise. The bias becomes more apparent when the conditions are divided into circular patterns and flapping, with $Z_{0,2}$ being negatively biased when the vortex shedding shows an anticlockwise circular pattern and positively biased when it shows a clockwise circular pattern, while the probability distribution of $Z_{0,2}$ in the case of flapping is the same as that of $Z_{0,2}$ at all times, centred around zero. The results above show that azimuthal shear mode tends to appear not when the vortex shedding shows flapping, but when the vortex shedding shows a circular pattern. 
Additionally, the rotational direction of the vortex shedding is found to determine the direction of the azimuthal shear. The rotational direction of the vortex shedding and the moments generated by the azimuthal velocity fluctuation field of mode $(m,n)=(0,2)$ are also discussed. The bias distributions in Figs.~\ref{fig:prob_0_2} (a--d) show that anticlockwise moments act when the vortex shedding position changes in the anticlockwise direction and clockwise moments act when the vortex shedding position changes in the clockwise direction, which indicates that the direction of the moments matches the change in the vortex shedding position. These relationships are further discussed in \S~\ref{sec:Relation}.

\begin{figure}
    \centering
    \includegraphics[width=13cm]{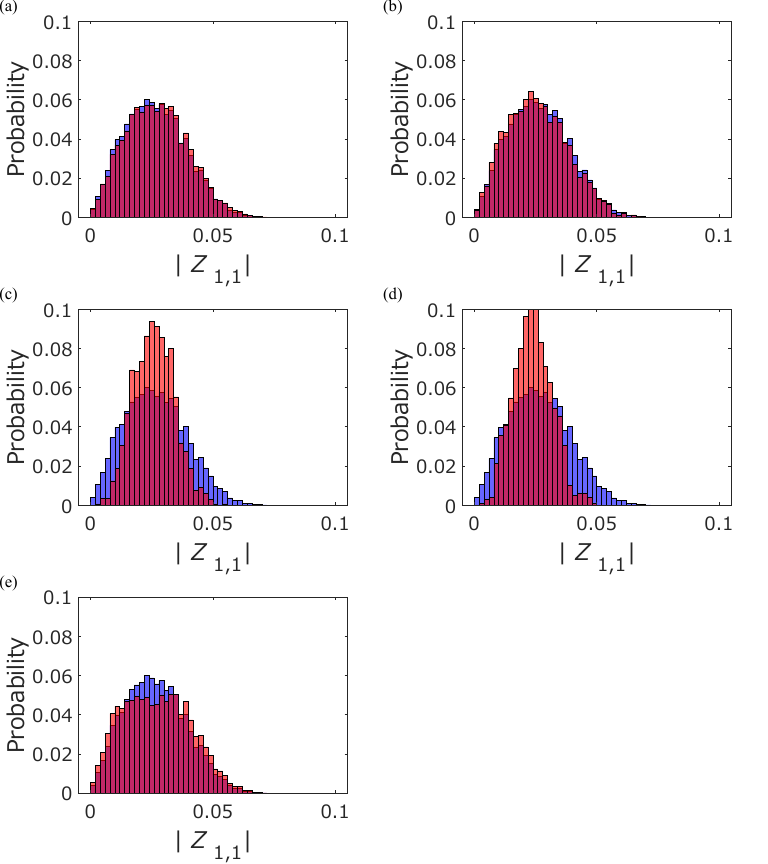}
    \caption{\label{fig:prob_amp} The probability distribution of $|Z_{1,1}|$ obtained by sampling under the condition that the state of vortex shedding is (a) anticlockwise, (b) clockwise, (c) anticlockwise circular, (d) clockwise circular and (e) flapping at $x/D=1.4$ in the case of $L/D=1.0$. The red and blue histograms show the results of the conditional sampling and the whole measurement time, respectively.}
\end{figure}

Figures~\ref{fig:taylor} (a, b) show that the magnitude of the fluctuations, i.e. $|Z_{1,1}|$, does not change significantly while the vortex shedding exhibits a circular pattern. Therefore, the same conditional sampling analyses for a circular pattern were conducted and a trend in $|Z_{1,1}|$ was investigated. Here, bandpass filtering of $0.1\le St\le0.2$ was applied to $Z_{1,1}$ in advance. Figures~\ref{fig:prob_amp} (a--e) show the results of sampling of $|Z_{1,1}|$ in the anticlockwise/clockwise, anticlockwise/clockwise circular and flapping patterns, respectively. The probability distribution for the whole measurement time, shown by the blue histogram, is similar to the probability distribution in Fig.~\ref{fig:prob_amp_ang} (a), and the maximum value is also found in the bin of $0.022\text{--}0.024$. Figures~\ref{fig:taylor} (c, d) show that $|Z_{1,1}|$ concentrates near the maximum value of the probability distribution for the whole measurement time in the case of the circular pattern. On the other hand, the probability is slightly higher outside that area in the case of flapping patterns. \cite{rigas2015diffusive} reported that the positional fluctuations of the centre of pressure (CoP) can be explained by the Mexican-hat-shaped potential field of the nonlinear Langevin model at the base of an axisymmetric bluff body with $Re_D=1.88\times10^5$, and the position in the $r$ direction where the CoP probability distribution is maximal corresponds to the position of the potential well. The probability distribution of $|Z_{1,1}|$ corresponding to the $r$-position of the vortex shedding shown in the blue histogram is similar to theirs, which implies that the vortex shedding position of the wake of the freestream-aligned circular cylinder can also be explained by a Mexican-hat-shaped potential field. In short, the $r$-position of the vortex shedding has small fluctuations and is stable near the potential well with an almost constant velocity in the azimuthal direction when the vortex shedding draws a circular pattern. In other words, since the vortex shedding position is trapped in the potential well and becomes stable, the $r$-position and angular velocity of the vortex shedding are considered to be almost constant. On the other hand, the vortex shedding position is considered to move through the centre of the potential field on a plane passing the cylinder axis (with a $\pi$ shift in azimuthal position) when the vortex shedding draws a flapping pattern. A transition to a flapping pattern is considered to occur when the $r$-position is more distant from the cylinder axis during the fluctuation of the $r$-position of the vortex shedding because it can cross through the potential well and over the potential mountain at the cylinder axis. The reason why the probability is lower than the whole measurement time near the potential well in the case of the flapping pattern shown in Fig.~\ref{fig:prob_amp} (e) is considered to be due to the uniform sampling of the measurements and the large velocity in the $r$ direction of the vortex shedding position near the potential well.

Although not shown in the paper, the change in the azimuthal shear mode or $|Z_{1,1}|$, associated with the vortex shedding pattern shown in Figs~\ref{fig:prob_0_2} and \ref{fig:prob_amp}, is also observed for $L/D=1.5$ and 2.0. Here, the analysis method is the same as described above. Therefore, it is expected that the shear layer reattachment does not change the flow characteristics due to the vortex shedding patterns.

\subsection{Relationship between characteristic fluctuations}\label{sec:Relation}
\begin{figure}
    \centering
    \includegraphics[width=12cm]{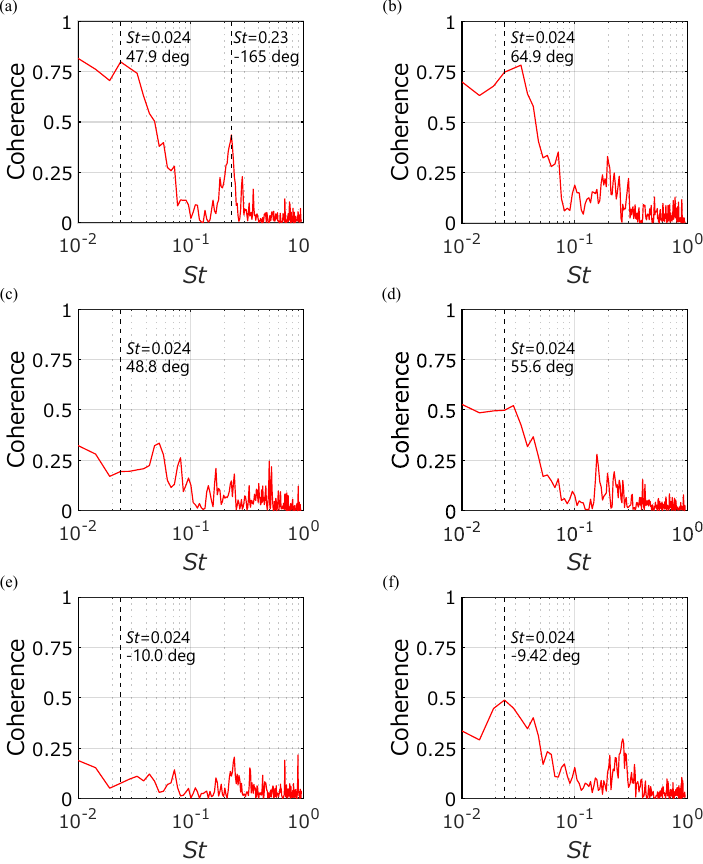}
    \caption{\label{fig:coh} The coherence between (a, b) $Z_{0,1}$--$|Z_{1,1}|$, (c, d) $Z_{0,1}$--$|Z_{2,1}|$ and (e, f) $|Z_{1,1}|$--$|Z_{2,1}|$ at (a, c, e) $x/D=1.0$ and (b, d, f) 2.0 in the case of $L/D=1.0$. The phase difference at the focused frequency is shown on side the dotted line.}
\end{figure}

Previous studies on the wake flow of a disc have discussed which characteristic fluid phenomena each mode corresponds to based on the results of modal decomposition. Similar discussions have been developed in \S~\ref{sec:Mode} and \ref{sec:VSposition} of this paper. The fluctuations of $Z_{0,1}$ and $|Z_{1,1}|$ in the case of the nonreattaching flow were found to be larger in the low-frequency region of $St\approx0.024$, while those of $Z_{0,1}$, $Z_{2,1}$ and $|Z_{1,1}|$ were larger at $St\approx0.23$. Coherence and phase differences were calculated and the relationships between the modes were investigated for $L/D=1.0$. The results shown in Fig.~\ref{fig:coh} illustrate the coherence between $Z_{0,1}$--$|Z_{1,1}|$, $Z_{0,1}$--$|Z_{2,1}|$ and $|Z_{1,1}|$--$|Z_{2,1}|$. Here, absolute values are taken for the mode $m\neq0$ because the mode coefficients are complex values. Another reason for the analysis was that the cylinder wake is statistically homogeneous in the azimuthal direction. Therefore, it was considered that correlations could be found in the absolute values of the mode coefficients corresponding to the amplitude of fluctuations and in quantities such as $|\mathrm{arg}~Z_{1,1}/dt|$ that are independent of the direction of rotation of the spatial pattern.

\begin{figure}
    \centering
    \includegraphics[width=13cm]{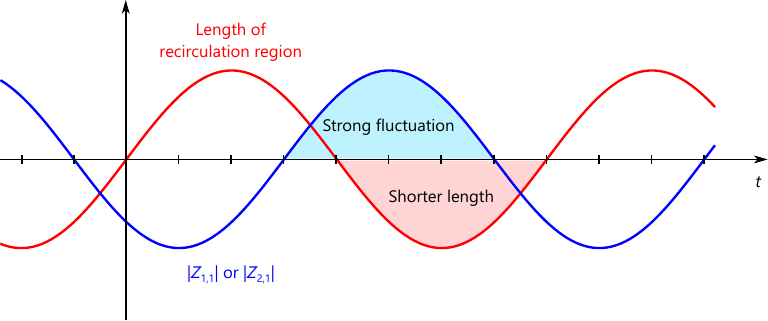}
    \caption{\label{fig:relation} The relationship among the length of the recirculation region, $|Z_{1,1}|$ and $|Z_{2,1}|$.}
\end{figure}

At first, high coherence is observed in the low-frequency region of $St\approx0.024$. The values are approximately 0.75 between $Z_{0,1}$--$|Z_{1,1}|$ as shown in Figs.~\ref{fig:coh} (a, b). The coherence between $Z_{0,1}$--$|Z_{2,1}|$ is relatively high downstream from $x/D=1.4$ near the downstream end of the recirculation region with a value of about 0.5 as shown in Figs.~\ref{fig:coh} (c, d), and the phase difference is similar to that between $Z_{0,1}$--$|Z_{1,1}|$. Here, the result of the coherence at $x/D=1.4$ is not shown. Moreover, relatively high coherence is observed downstream of the recirculation region. The low-frequency fluctuations for $St\le0.05$ in $Z_{2,1}$ are larger downstream of the recirculation region, as shown in Fig.~\ref{fig:PSD_Z} (d). This low-frequency fluctuation is considered to be due to streaks, discussed in \S~\ref{sec:Mode}. In contrast, fluctuations at $St\approx0.23$ due to the double-helix structure were also observed at $x/D=1.0$ and 2.0, which corresponds to inside and outside the recirculation region, respectively. For these reasons, the low-frequency fluctuations in $|Z_{2,1}|$, which are correlated with $Z_{0,1}$, are considered to be mainly related to streaks. A similar trend is also observed between $|Z_{1,1}|$--$|Z_{2,1}|$ as shown in Fig.~\ref{fig:coh} (e, f). Figure~\ref{fig:relation} summarises the relationship between the length of the recirculation region and the strength of fluctuations in mode $(m,n)=(1,1),(2,1)$ based on the phase difference between $Z_{0,1}$--$|Z_{1,1}|$ and $Z_{0,1}$--$|Z_{2,1}|$. The fluctuations in mode $(m,n)=(1,1),(2,1)$ are strong when the recirculation region is shorter than the mean field due to bubble pumping (red region in the figure). However, the causality between these modes is not clear, and therefore the transfer entropy between $Z_{0,1}$, $|Z_{1,1}|$ and $|Z_{2,1}|$ was calculated. 

\begin{figure}
    \centering
    \includegraphics[width=13cm]{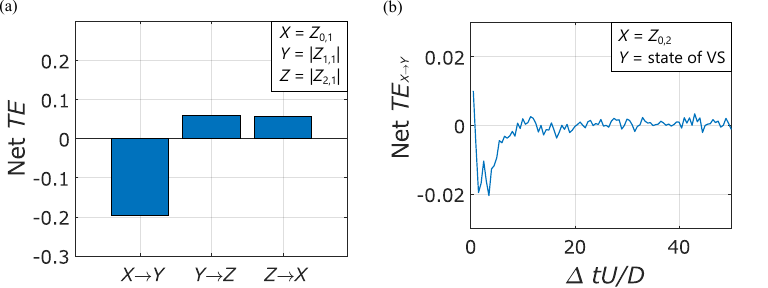}
    \caption{\label{fig:NTE} The net transfer entropy for (a) three different paths between $Z_{0,1}$, $|Z_{1,1}|$ and $|Z_{2,1}|$ and (b) a path between the $Z_{0,2}$ and the vortex shedding states.}
\end{figure}

Entropy $H$ in information theory is calculated by the following equation:
\begin{eqnarray}
    \label{eq:entropy}
    H(X) &=& -\sum_{x \in X} p(x)\log_2p(x),
\end{eqnarray}
where $p$ is the probability density function with $x$ as the random variable. If two or more random variables are involved, it is called joint entropy and is calculated as follows:
\begin{eqnarray}
    \label{eq:CE}
    H(X,Y) &=& -\sum_{x \in X}\sum_{y \in Y} p(x,y)\log_2p(x,y).
\end{eqnarray}
The probability distribution in the present study was created by discretizing the data in six bins between the maximum and minimum values of the data and dividing the number of samples in each bin by the total number of sample points. The analysis for setting the number of bins is described in Appendix~\ref{sec:Bin}. The transfer entropy $TE$ is calculated by Eq.~\ref{eq:TE}:
\begin{eqnarray}
    \label{eq:TE}
    \begin{split}
    TE_{X \rightarrow Y} =&~H(Y(t)|Y(t-\Delta t))-H(Y(t)|Y(t-\Delta t),X(t-\Delta t))\\
                         =&~H(Y(t),Y(t-\Delta t))-H(Y(t-\Delta t))\\
                         &-H(Y(t),Y(t-\Delta t),X(t-\Delta t))+H(Y(t-\Delta t),X(t-\Delta t)),
    \end{split}
\end{eqnarray}
where $\Delta t$ is the time lag and was set to 0.03~s ($\Delta tU/D=6.3$) in the calculation of transfer entropy between $Z_{0,1}$, $|Z_{1,1}|$ and $|Z_{2,1}|$. The spurious transfer entropy caused by statistical errors is removed by the following equation \citep{lozano2020causality, milani2020shear}:
\begin{eqnarray}
    \Tilde{TE}_{X \rightarrow Y} &=& TE_{X \rightarrow Y}-TE_{X^{\rm shuffled} \rightarrow Y},
\end{eqnarray}
where $X^{\rm shuffled}$ is the data with $X$ randomly permuted. and the dominant direction of information flow is determined by the net transfer entropy $TE_{\rm net}$ calculated by Eq.~\ref{eq:NTE}:
\begin{eqnarray}
    \label{eq:NTE}
    TE_{{\rm net},X \rightarrow Y} &=& \Tilde{TE}_{X \rightarrow Y}-\Tilde{TE}_{Y \rightarrow X}.
\end{eqnarray}
Positive $TE_{{\rm net},X \rightarrow Y}$ indicates that the information flow is dominant in the $X$ to $Y$ direction. In the present study, the data for 24.7~s at $x/D=1.4$ was used to create a discrete probability distribution, and entropy and joint entropy were calculated to compute $TE_{\rm net}$. Figure~\ref{fig:NTE} (a) shows the results of the net transfer entropy for three different paths between $Z_{0,1}$, $|Z_{1,1}|$ and $|Z_{2,1}|$. Here, $Z_{0,1}$ and $Z_{2,1}$ were lowpass filtered with $St\le0.05$ and $Z_{1,1}$ was bandpass filtered with $0.1\le St\le0.2$, respectively, so as to focus on the recirculation bubble pumping, the large-scale vortex shedding and the streaks. Furthermore, calculated $|Z_{1,1}|$ and $|Z_{2,1}|$ were processed with a lowpass filtering of $St\le0.05$. The $TE_{\rm net}$ between $Z_{0,1}$--$|Z_{1,1}|$ is \add{negative}, which indicates that the information flow is dominant in the direction from \add{the amplitude of mode $(m,n)=(1,1)$, which is considered to correspond to the radial position of the vortex shedding position, to bubble pumping represented by mode $(m,n)=(0,1)$}.  Similarly, the direction \add{from amplitude of mode $(m,n)=(2,1)$ to mode $(m,n)=(0,1)$} is dominant between $Z_{0,1}$--$|Z_{2,1}|$. The dominant flow of information between $|Z_{1,1}|$--$|Z_{2,1}|$ is from the amplitude of mode $(m,n)=(1,1)$ to the amplitude of mode $(m,n)=(2,1)$, \add{and $TE_{\rm net}$ is comparable to it between $Z_{0,1}$--$|Z_{2,1}|$}. Information is mainly transferred from the one with a low wavenumber to the one with a high wavenumber \add{only in the $|Z_{1,1}|$--$|Z_{2,1}|$ case}. In short, \add{the length of the recirculation region is considered to be affected by changes in the radial position of the vortex shedding and the strength of the streak. The results in Figs.~\ref{fig:coh} and \ref{fig:NTE} (a) indicate that the large-scale vortex shedding especially has a significant effect on the bubble pumping. Stronger mixing by flapping or helical structures consequently leads to a shorter recirculation region when flapping or helical structures exhibiting circular patterns of large-scale vortex shedding are stronger, while conversely, weaker mixing by flapping or helical structures results in a longer recirculation region. Hence, the results implicate that the low-frequency fluctuations in large-scale vortex shedding strength are mainly producing the bubble pumping.}


Secondly, relatively high coherence between $Z_{0,1}$--$|Z_{1,1}|$ at $St=0.23$, which is apparent in the $x$-position where the recirculation region is located, as shown in Fig.~\ref{fig:coh}. The relationship between $Z_{0,1}$--$|Z_{1,1}|$ at this frequency implies that the vortex shedding position moves in the $r$-positive direction as the length of the recirculation region increases, which is the reversal of the relationship in the low-frequency region. High-frequency fluctuations in the azimuthal mode $m=0$ have been observed in previous studies \citep{berger1990coherent, fuchs1979large, nidhan2020spectral, nekkanti2023large} but neither they have not been discussed in detail, nor the present study does not provide a three-dimensional velocity field. Thus, the discussion of physical relationships is difficult. However, the fact that the relationship is found only in the recirculation region is consistent with the finding by \cite{meliga2009global} that the nonlinear interaction between unstable modes occurs only in the recirculation region.

Another feature found only within the recirculation region is the rotational flow reported by \cite{zhang2023coherent}. They suggested that the planar symmetric vortex loops formed behind an axisymmetric body are collapsed by the azimuthal shear mode that appears in the recirculation region, which results in a twisted form of vortex loops known as the ``Yin-Yang'' pattern. Mode $(m,n)=(0,2)$, which represents azimuthal shear mode, is also found in the present study, as shown in the figure, and the results in the section reveal that mode $(m,n)=(0,2)$ is associated with the state of vortex shedding. However, the causality between them is not clear. For this reason, a causality investigation was carried out using transfer entropy in the same way as described above. Since there are three vortex shedding patterns in the calculation, the number of bins for the vortex shedding state is three, and the number of bins for $Z_{0,2}$ is six. Figure~\ref{fig:NTE} (b) shows the $TE_{\rm net}$ between the $Z_{0,2}$ and vortex shedding states and also represents the interesting profile in the range with small $\Delta t$. The absolute value of $TE_{\rm net}$ reaches a maximum at \add{$\Delta tU/D=3.5$}, and the dominant direction is \add{from the vortex shedding state to $Z_{0,2}$}. However, the information flow in the opposite direction becomes dominant with increasing $\Delta t$. \cite{zhang2023coherent} suggest that a plane-symmetric vortex is twisted by azimuthal shear mode and an RSB distribution of vorticity appears, and since the present study also shows that the change in vortex shedding position coincides with the direction of the moment in the cross-section, it is reasonable to assume that the azimuthal shear mode determines the vortex shedding state. In contrast, changes in the vortex shedding position are not expected to generate azimuthal shear or moments. Two vortices with opposite signs of streamwise vorticity in the $yz$ plane do not form an RSB distribution at each other's induced velocities and have zero moments in the cross-section. \add{The relationship between this vortex shedding pattern and azimuthal shear mode remains to be further investigated}. 

The above discussion shows that axisymmetric fluctuations, such as bubble pumping, are related to the radial position of vortex shedding and the strength of the streak, especially in the low-frequency range. \add{Specifically, changes in vortex shedding have been suggested to cause the bubble pumping}. The large-scale vortex shedding is linked to aerodynamic forces acting in the lateral, pitch and yaw directions of the cylinder, while fluctuations due to bubble pumping are closely related to base pressure \citep{yokota2021analysis, kuwata2021flow, nonomura2018effect, shinji2020aerodynamic}. This suggests the possibility of predicting the aerodynamic forces acting in the lateral and rotational directions by measuring base pressure in axisymmetric bodies such as those focused on in the present study. However, since the causes of azimuthal shear mode are not known and could not be clarified in the present study, further investigation is required for this point.

\section{Conclusions}\label{sec:Conclusion}
The present study has focused on the three-dimensional large-scale wake structure of characteristic phenomena appearing behind the freestream-aligned circular cylinder. Particularly, the relationship between the modes of velocity fluctuations in the nonreattaching flow, the change in the state of the large-scale vortex shedding, and the relationship between axisymmetric fluctuations such as the recirculation bubble pumping and the azimuthal shear mode have been clarified. Experimental investigations in a flow-interference-free condition were carried out by contactless two-dimensional three-component velocity measurements, which combined the 0.3-m MSBS with the stereo PIV measurement system. Three cylindrical models with $L/D=1.0$, 1.5 and 2.0 were used for the wind tunnel tests. Firstly, the results of the flow properties show that the errors are large for the time-averaged field compared to previous studies, but there is a good agreement for the fluctuation field. The modal decomposition, which combines azimuthal Fourier decomposition and proper orthogonal decomposition, is employed for the field of velocity fluctuations, and the relationship between modes and vortex shedding has been discussed.

The eigenvalue spectra show that the $m=1$ mode is dominant regardless of $L/D$ and $x$-position in the range of the present study. The order of contribution for higher wavenumber modes than $m=1$ is the order of increasing wavenumber, but the order of contribution for $m=0$ depends on the $L/D$ and $x$-position. The eigenfunctions and the PSDs of the real part of the mode coefficients are subsequently presented, which focus on the case of $L/D=1.0$ where characteristic structures appear. The eigenfunctions of mode $(m,n)=(0,1)$ are dominated by the velocity fluctuations in the $x$ direction and also produce the in-phase velocity fluctuations in the $r$ direction. Since the fluctuations of this mode at $St\approx0.024$ are large at $x$-positions corresponding to the inside of the recirculation region, these fluctuations were treated as the recirculation bubble pumping. On the other hand, the fluctuations in this mode were also found to be large at $St\approx0.23$, but it could not be clarified what kind of fluid phenomena correspond to these fluctuations. The mode $(m,n)=(0,2)$ shows that the main component is the $\theta$ component and the eigenfunctions represent a velocity distribution that causes azimuthal shear mode. However, no characteristic trend is observed from the PSDs. The eigenfunctions of the dominant mode $(m,n)=(1,1)$ illustrate the acceleration and deceleration regions of $u_x$ and two vortices with opposite signs of vorticity in the freestream direction. A clear peak is observed at $St=0.129$ in this mode, and this fluctuation corresponds to large-scale vortex shedding. The eigenfunctions for mode $(m,n)=(2,1)$ exhibit four vortices with vorticity in the freestream direction with switching signs alternately in the azimuthal direction. The fluctuations at $St\approx0.23$ of this mode are considered to correspond to a double-helix structure, and the low-frequency fluctuations are considered to correspond to a streak. 

One of the relationships between characteristic phenomena is the relationship between the shedding position of the large-scale vortex structure and the recirculation bubble pumping, as suggested by \cite{yang2015low}. In the present study, the relationship is discussed mainly based on the mode $(m,n)=(1,1)$. The vortex shedding showed circular patterns in anticlockwise and clockwise directions, flapping patterns, and a mixture of them. The state is irregularly changed with time. Amplitude changes in the mode coefficients corresponding to $r$-position fluctuations of the vortex shedding occur at $St\approx0.26$ as a doubling of the fluctuation frequency due to the large-scale vortex shedding. This amplitude change is considered to be due to the vortex shedding in a flapping pattern. The association between these three vortex shedding patterns and the recirculation bubble pumping was investigated by conditional sampling, but no association was found between them. However, a link was found between the vortex shedding pattern and the mode $(m,n)=(0,2)$, which indicates azimuthal shear mode. Azimuthal shear mode tends to appear when vortex shedding exhibits a circular pattern, and the direction of shear also tends to vary depending on whether the circular direction is anticlockwise or clockwise. Similarly, a trend occurred in the moments calculated from the field of azimuthal velocity fluctuations. The direction of this moment coincides with the direction of the vortex shedding position and other observations suggest that the azimuthal shear mode is related to the circular pattern of vortex shedding. Furthermore, the relationship between the vortex shedding pattern and the $r$-position of the vortex shedding was also observed. The $r$-position of the vortex shedding is stable with an almost constant azimuthal velocity when the vortex shedding exhibits a circular pattern.

A relationship was observed between modes other than the vortex shedding patterns, and the coherence is high between $Z_{0,1}$--$|Z_{1,1}|$, $Z_{0,1}$--$|Z_{2,1}|$, and $|Z_{1,1}|$--$|Z_{2,1}|$. In the low-frequency region of $St\approx0.024$, the phase difference indicates that $Z_{1,1}$ and $Z_{2,1}$ become larger when the recirculation region is shorter due to the recirculation bubble pumping, i.e. the fluctuations of the vortex shedding position in the $r$ direction become larger and the streaks are stronger. \add{The vortex shedding and streaks were found to affect the state of the recirculation region}. In the high-frequency region of $St=0.23$, the coherence is relatively high between $Z_{0,1}$--$|Z_{1,1}|$, and the phase difference indicates that the vortex shedding position moves in the positive direction of $r$ as the length of the recirculation region increases. This relationship is only observed within the recirculation region, which is in agreement with the finding of \cite{meliga2009global} that the interactions between modes only occur within the recirculation region.

The present study has partly clarified the relationships between the characteristic flow structures that appear in a nonreattaching flow. The relationship between the bubble pumping and the large-scale vortex shedding could be used to predict the aerodynamic forces acting on a body and control them with a plasma actuator \citep{aono2019mechanisms}, etc. Further clarification of the relationship is desired in the future. Further understanding could be improved by experimental investigations, such as pressure fluctuation field measurements at the base and sides of the cylinder, and simultaneous dual-plane PIV measurements in planes parallel and perpendicular to the cylinder axis. Since it is rare for the flow to come from the front in the applications listed in the introduction, similar investigations to the present study are required for the case where the cylinders are angled to the flow, which is closer to the application.


\backsection[Acknowledgements]{We thank Drs. K. Asai and Y. Ozawa who gave ideas to construct the stereo PIV measurement system compatible with the MSBS.}

\backsection[Funding]{This work was supported by JSPS KAKENHI (grant numbers 18H03809, 21H04586, 21J20673, 22KJ0175).}

\backsection[Declaration of interests]{. The authors report no conflict of interest.}

\backsection[Author ORCID]{\\S. Yokota, https://orcid.org/0000-0002-0004-7015;\\ 
T. Nonomura, https://orcid.org/0000-0001-7739-7104}

\backsection[Author contributions]{S.Y. conducted the experiment and analysed data and T.N. proposed the method of the analysis and supervised the study. All authors contributed equally to reaching conclusions and in writing the paper.}

\appendix



\section{Freestream velocity and its correction}\label{sec:FSVel}
Wind tunnel calibration tests were performed in the present study because the test section with different dimensions from the normal one was used and the profiles of velocity and turbulence intensity were expected to be different. The tests were performed to simulate the conditions of stereo PIV measurements, although particle introduction was not conducted. The profiles of the velocity and the turbulence intensity were measured at the centre of the test section in the $y$ direction with traversing in the $z$ direction at 750 and 874.5~mm from the upstream end of the test section, respectively. 

\begin{figure}
    \centering
    \includegraphics[width=13cm]{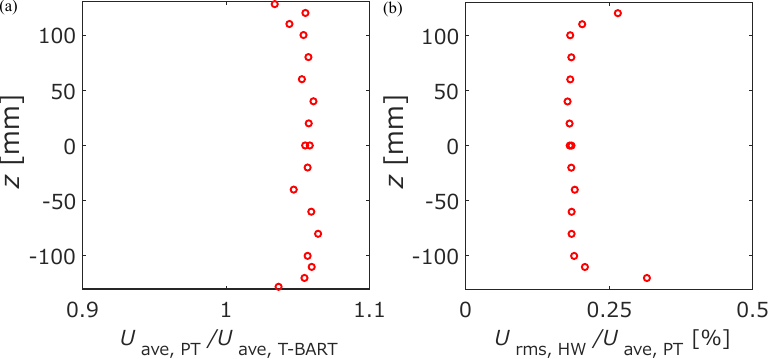}
    \caption{\label{fig:vel_cor} (a) The time-averaged velocity and (b) turbulence intensity profiles in the wind tunnel calibration tests}
\end{figure}

The velocity profile was measured using an L-shaped pitot tube. The pitot coefficient of this pitot tube is 1. Figure~\ref{fig:vel_cor} (a) shows the results of the velocity profile measurements. $U_{\rm ave, T-BART}$ are freestream velocities obtained by the wind tunnel data measurement system and calculated from the differential pressure before and after the contraction part. The test section used in the present study has a narrower width in the $z$ direction than normal, which accelerates the flow. Since $U_{\rm ave, T-BART}$ was obtained in the experiment, a correction factor was multiplied to it to obtain the freestream velocity. The correction factor was 1.05, which is the average value for the range -100~mm$\le z\le$100~mm.

The turbulence intensity was measured by a hot-wire anemometer. The used devices were a hot-wire anemometer (CTA-002, Institute of Flow Research), a probe (type 55P11, Dantec Dynamics), a low-pass filter (PGF-8ELA, Japan Audio), a filter (FV-665, NF Electronic Instruments), an amplifier (3628, NF Electronic Instruments) and a recording device (USB-6363, National Instruments). Figure~\ref{fig:vel_cor} (b) shows the results of the turbulence intensity measurements. Turbulence due to the boundary layer was strong near the test section wall, but the average value for the range -100~mm$\le z\le$100~mm mm was low (0.18\%), which shows that the effect of change in dimension was small.

The freestream velocity obtained by the wind tunnel data measurement system was set to 10~m/s in the present study. The empirical findings of our group show that the fluctuations of the freestream velocity increase when it is set below 10~m/s because T-BART is a suction-type wind tunnel and the outlet is connected to outside building in the case of PIV measurements. Another reason is that the positional fluctuations of the levitated model during wind-on conditions are smaller when the flow velocity is lower. The above freestream velocity settings and the flow acceleration in the test section resulted in a slightly different Reynolds number ($3.46\times10^4$) condition from the previous studies \citep{yokota2021analysis, yokota2022instability, yokota2023effect}.

\section{Sensitivity of the number of bins on transfer entropy}\label{sec:Bin}
The calculation of transfer entropy requires a probability density function, and the probability distribution was created by discretising the mode coefficients or their absolute values, as described in \S~\ref{sec:Relation}. The number of bins affects the magnitude of the transfer entropy \citep{milani2020shear}. Therefore, the effect of the number of bins on transfer entropy was analysed and $\Delta t$ was set to obtain the results used in the discussion.

\begin{figure}
    \centering
    \includegraphics[width=13cm]{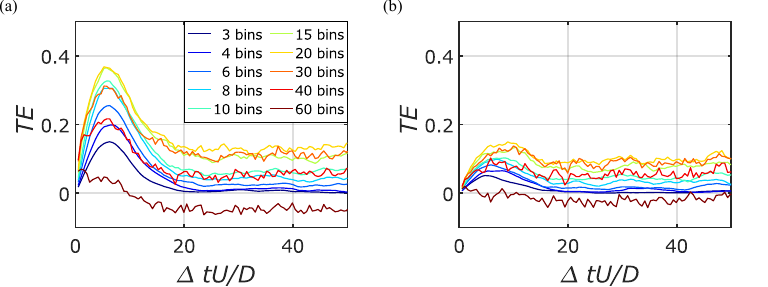}
    \caption{\label{fig:TE_para} Profiles of transfer entropy between (a) $|Z_{1,1}|\rightarrow Z_{0,1}$ and (b) $|Z_{2,1}|\rightarrow Z_{0,1}$ with respect to time lag $\Delta t$}
\end{figure}

Figures~\ref{fig:TE_para} (a, b) show the transfer entropy $\Tilde{TE}$ for different numbers of bins for $|Z_{1,1}|\rightarrow Z_{0,1}$ and $|Z_{2,1}|\rightarrow Z_{0,1}$, respectively. The horizontal axis is a time lag $\Delta t$ nondimensionalised by the freestream velocity $U$ and the cylinder diameter $D$. A smooth profile is apparent near the peak when the number of bins is three to ten, but the profile becomes less smooth when the number of bins exceeds 15, and the peak is no longer visible at bin number of 60. Although the number of bins is eight and ten, which represent smooth profiles, the transfer entropy becomes a non-zero value when $\Delta t$ is increased. The transfer entropy should be almost zero as the time lag is larger because there is no relationship with the state at time $t$. The number of bins in the analysis was set to 6 because they show a clear peak, the profile is smooth and the transfer entropy is almost zero at large $\Delta t$.

\begin{figure}
    \centering
    \includegraphics[width=13cm]{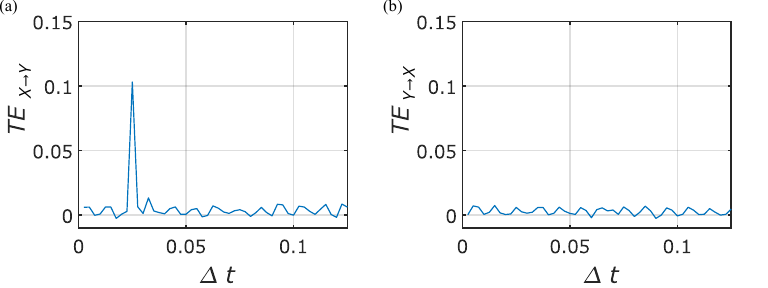}
    \caption{\label{fig:TE_test} Profiles of transfer entropy between (a) $X(x_n)\rightarrow Y(y_n)$ and (b) $Y(y_n)\rightarrow X(x_n)$ with respect to time lag $\Delta t$}
\end{figure}

\add{A test was also conducted to confirm that the analysis code used in the present study could correctly evaluate the direction of the information flow, based on the model expressed in the following equations:
\begin{eqnarray}
    \label{eq:TE_test}
    x_n &=& \sin(2\pi ft)+\epsilon_{x_n}, \\
    y_n &=& -0.5y_{n-1}+0.5x_{n-10}+\epsilon_{y_n},
\end{eqnarray}
where $f=50$~Hz, $t=0\text{--}25$~s ($\Delta t=0.0025$~s), $y_1=1$, and $\epsilon$ is white Gaussian noise. The number of bins between the maximum and minimum values for each dataset was set to 6, and probability distributions were created from samples of 9991 points. Figures~\ref{fig:TE_test} (a, b) show the transfer entropy for $X(x_n)\rightarrow Y(y_n)$ and $Y(y_n)\rightarrow X(x_n)$, respectively, and indicate high transfer entropy in the correct direction and time lag.}

\begin{figure}
    \centering
    \includegraphics[width=13cm]{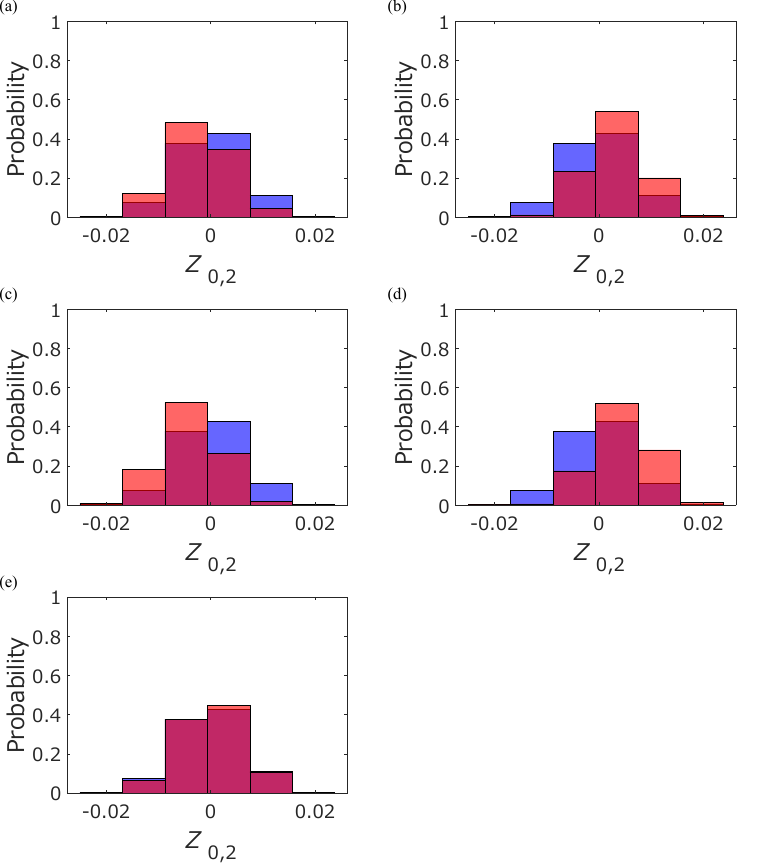}
    \caption{\label{fig:prob_0_2_6} The probability distribution of $Z_{0,2}$ obtained by sampling under the condition that the state of vortex shedding is (a) anticlockwise, (b) clockwise, (c) anticlockwise circular, (d) clockwise circular and (e) flapping at $x/D=1.4$ in the case with $L/D=1.0$ and the number of bins of 6. The red and blue histograms show the results of the conditional sampling and the whole measurement time, respectively.}
\end{figure}

The transfer entropy between the vortex shedding state and $Z_{0,2}$ was also calculated with a bin number of six. On the other hand, the conditional sampling results in Fig.~\ref{fig:prob_0_2} are shown with a higher number of bins than this. Figure~\ref{fig:prob_0_2_6} shows the conditional sampling results when the number of bins is set to 6. Although there was an increase in probability due to the reduced number of bins, the qualitative trends were not changed.

\bibliographystyle{jfm}
\bibliography{main.bib}


\end{document}